\documentclass[twocolumn,showpacs,prd,floatfix,axodraw]{revtex4}
\usepackage{mathrsfs}
\usepackage{graphicx,,booktabs,bm}
\usepackage{overpic}
\usepackage{color}
\usepackage{amssymb}

\usepackage{mathrsfs,bm,amsmath,amssymb}
\usepackage{longtable,lscape}
\usepackage{txfonts}
\usepackage{amssymb}
\usepackage{indentfirst}
\usepackage{graphicx,,booktabs}
\usepackage{multirow,ulem}

\usepackage{color}
\usepackage{amssymb}

\definecolor{cover}{rgb}{0.77,0.87,0.88}
\definecolor{blueone}{rgb}{0.1,0.1,.7}
\definecolor{citec}{rgb}{0.14,0.47,0.09}
\definecolor{two}{rgb}{0.0,0.5,0.}
\definecolor{three}{rgb}{.5,.1,0.15}
\usepackage[bookmarks=true,bookmarksopen=false,plainpages=false,breaklinks=true,
   bookmarksnumbered=true,hypertexnames=false,
   filecolor=blue,urlcolor=three,menucolor=three,
   linkcolor=three,citecolor=blueone, colorlinks,
   anchorcolor=blue,runcolor=pink,frenchlinks=red
   pdfstartview=FitH,pdftitle=title,%
   pdfauthor=author]{hyperref}

\begin{document}
\title{Probing the Molecular Nature of $D_{sJ}(2860)$ and $D_{s3}(2860)$ through Their Decays}
\author{Qing Lu}
\affiliation{Key Laboratory of Computational Physics of Sichuan Province, School of Mathematics and Physics, Yibin University, Yibin 644000, China}
\author{Dan Dan Feng}
\affiliation{School of Mathematics, Physics and Statistics, Sichuan Minzu College, Kangding 626001, China}
\author{Cai Cheng}
\affiliation{School of Physics and Electronic Engineering, Sichuan Normal University, Chengdu 610101, China}
\author{Yin Huang\footnote{corresponding author}} \email{huangy2019@swjtu.edu.cn}
\affiliation{School of Physical Science and Technology, Southwest Jiaotong University, Chengdu 610031,China}

\begin{abstract}
In recent years, advances in experimental techniques have led to the discovery of an increasing number of exotic hadronic states, among which hadronic molecular states are widely regarded as an important avenue for understanding their internal structures. In particular, high-partial-wave molecular configurations, in which the constituents carry nonzero relative orbital angular momentum $l$, especially $P$-wave molecules with $l=1$, are currently attracting considerable attention. In this work, we investigate the molecular interpretations of $D_{sJ}(2860)$ and $D_{s3}(2860)$ by calculating their strong decay widths within the effective Lagrangian approach.
For $D_{sJ}(2860)$, we consider two possible molecular configurations, namely, an $S$-wave $D_1K$ molecule and a $P$-wave $D^*K^*$ molecule, while treating $D_{s3}(2860)$ as a $P$-wave $D^*K^*$ molecule. By fitting the LHCb-measured $\cos\theta(\bar{D}^0K^-)$ angular distribution, which receives contributions only from $D_{sJ}(2860)$ and $D_{s3}(2860)$, and their measured decay widths in the $B_s^0\to\bar{D}^0K^-\pi^+$ decay, we determine the corresponding molecular coupling constants and four-momentum cutoff parameter $\Lambda$.  Using the resulting parameters, we calculate the total decay widths of the two states and the ratios $R$ of their partial decay widths into the $D^*K$ and $DK$ final states. Our results show that, for $D_{sJ}(2860)$, the $S$-wave $D_1K$ molecular interpretation yields a total decay width consistent with the experimental data throughout the fitted parameter range and provides a slightly better description than the $P$-wave $D^*K^*$ interpretation. However, when $D_{sJ}(2860)$ and $D_{s3}(2860)$ are interpreted as $D_1K$ and $D^*K^*$ molecular states, respectively, all predicted values of $R_{D_{s3}(2860)}$ lie above the experimental upper limit, disfavoring this combined molecular assignment. In contrast, if both states are interpreted as $P$-wave $D^*K^*$ molecules, the predicted $R_{D_{s3}(2860)}$ agrees well with the experimental measurement over the range $\Lambda=1.51$--$1.73$ GeV. This assignment also yields a prediction of approximately $1.26$--$1.29$ for $R_{D_{sJ}(2860)}$, which has not yet been measured experimentally. Considering both the total decay widths and the ratios of partial decay widths, our results support the interpretation of $D_{sJ}(2860)$ and $D_{s3}(2860)$ as $P$-wave $D^*K^*$ molecular states. Future measurements of $R_{D_{sJ}(2860)}$ could provide an additional test of this molecular structure.
\end{abstract}

\date{\today}

%\pacs{{13.60.Le}{inner structures}   \and {13.85.Lg}{Schr\"{o}dinger equation} \and  {25.30.-c}{molecular state}  }}

\maketitle
\section{Introduction}\label{sec:intro}
In recent years, with the continuous advancement of experimental techniques, an increasing number of exotic hadronic states have been discovered~\cite{ParticleDataGroup:2026mpi}.
Exotic hadrons refer to states with internal structures more complicated than those described by the conventional quark model, in which mesons are generally composed of a quark-antiquark
pair while baryons consist of three quarks~\cite{Godfrey:1985xj,Koniuk:1979vy,Isgur:1978wd}.  For instance, the $X(3872)$, discovered in 2003~\cite{Belle:2003nnu}, exhibits decay
properties that are difficult to accommodate within the conventional quark model. In addition, the hidden-charm pentaquark states $P_c$ and $P_{cs}$, which contain a $c\bar{c}$ pair,
have also been observed experimentally~\cite{LHCb:2015yax,LHCb:2016ztz,LHCb:2016lve,LHCb:2019kea,LHCb:2020jpq,LHCb:2022ogu}.  More recently, the $X(2370)$ has been observed
experimentally~\cite{BESIII:2026rzt} and is considered a promising glueball candidate.

Hadronic molecular states, which feature a hadronic structure analogous to that of the deuteron, a bound state of a proton and a neutron, are currently regarded as
one of the leading approaches to understanding the internal structures of exotic hadrons.  The aforementioned $X(3872)$, whose properties cannot be adequately explained
within the conventional quark model, is now widely regarded as a $\bar{D}D^{*}$ molecular state~\cite{Brambilla:2019esw,Chen:2022asf,Meng:2022ozq}.
Similarly, the observed hidden-charm pentaquark states can be interpreted as $\bar{D}^{(*)}\Sigma_c^{(*)}$ or $\bar{D}^{(*)}\Xi_c$ molecular states~\cite{Chen:2019bip,Guo:2019fdo,Xiao:2019aya,He:2019ify,Xiao:2019mvs,Roca:2015dva,Chen:2015moa,Chen:2015loa,Yang:2015bmv,Huang:2015uda,Du:2019pij}.
In addition, two well-known candidates for hadronic molecular states, $D_{s0}(2317)$ and $D_{s1}(2460)$, have been observed experimentally~\cite{BaBar:2003oey,
CLEO:2003ggt,Belle:2003kup,Belle:2003guh,BaBar:2004yux} and interpreted as $KD$ and $KD^{*}$ molecular states~\cite{Meng:2022ozq,Xie:2010zza,Guo:2006fu,Guo:2006rp,Gamermann:2006nm,Zhu:2019vnr,Mohler:2013rwa,Altenbuchinger:2013vwa},
respectively.

Although only 11 mesons containing a charm quark and a strange quark have been observed experimentally to date~\cite{ParticleDataGroup:2026mpi}, in addition to
$D_{s0}(2317)$ and $D_{s1}(2460)$, which can be interpreted as hadronic molecular states, there are also other candidates that may be understood as molecular states.
The $D_{s1}(2700)$ has been suggested to contain a substantial $DK^{*}$ molecular component, with one calculation estimating its probability to be as large as
approximately $51.3\%$~\cite{Hao:2022vwt}. The $D_{sJ}(2860)$ has also been interpreted as a $K^{*}D^{*}$~\cite{Hao:2022vwt,Jiang:2026rpu} or $D_1(2420)K$~\cite{Guo:2017jvc,
Wang:2025jcq,Guo:2011dd} hadronic molecular state, while the $D_{s3}(2860)$ may likewise contain a sizable $K^{*}D^{*}$ molecular component~\cite{Hao:2022vwt,Jiang:2026rpu}.
The broad $D_{sJ}(3040)$ has also been proposed as a candidate for a hadronic molecular state~\cite{Hao:2022vwt,Guo:2011dd}. It should be noted that, in Ref.~\cite{Guo:2011dd},
the interpretation of $D_{sJ}(2860)$ as a $D_1(2420)K$ molecular state was partly motivated by the calculated ratio
\begin{align}
R_{D_{sJ}(2860)}=
\frac{\Gamma(D_{sJ}(2860)\rightarrow D^*K)}
{\Gamma(D_{sJ}(2860)\rightarrow DK)}
=1.23\label{eq1-1},
\end{align}
which was found to be in good agreement with the experimental value $1.1\pm0.24$. However, according to the latest Particle Data Group (PDG) compilation, the measured
ratio $1.1\pm0.15\pm0.19$ is associated with the $D_{s3}(2860)$ state with $J^P=3^-$, rather than the $D_{s1}(2860)$ state.

In the absence of compelling evidence for their molecular nature, the conventional charmed-strange quark-antiquark structure of these states remains widely accepted,
while their specific spectroscopic assignments have yet to be firmly established and vary among different approaches~\cite{Wang:2014jua,Li:2010vx,Close:2006gr,Godfrey:2013aaa,Chen:2009zt,
Wang:2016hkf,Zhang:2006yj,Li:2009qu,Jiang:2024ftx}.  For instance, a systematic study of the spectroscopy and strong decays within a constituent quark model suggests that
the $D_{s1}(2710)$ is a $2^3S_1$--$1^3D_1$ mixed state, the $D_{sJ}(2860)$ may correspond to either the orthogonal partner of the $D_{s1}(2710)$ or the $D_s(1^3D_3)$ state,
and the $D_{sJ}(3040)$ can be naturally interpreted as either the $D_{s1}(2P)$ or $D^{'}_{s1}(2P)$ state~\cite{Li:2010vx}. These assignments for the $D_{s1}(2710)$ and
$D_{sJ}(2860)$ are also supported by strong-decay calculations within the $^3P_0$ model~\cite{Li:2009qu}; however, another $^3P_0$ model analysis~\cite{Zhang:2006yj}
yields decay widths that are inconsistent with these assignments.  More strikingly, in a semiclassical flux-tube model, the $D_{s1}(2700)$ is favored as the first
radially excited $1^-$ $D_s$ meson, the $D_{sJ}(2860)$ is identified as the orbitally excited $1^3D_3$ state with $J^P=3^-$, and the $D_{sJ}(3040)$ is interpreted as
the first radially excited $1^+$ $D_s$ meson~\cite{Chen:2009zt}.

To date, the decay widths of the $D_{s1}(2860)$ and $D_{s3}(2860)$ have not yet been systematically investigated within the hadronic molecular picture. In this work, we employ
the effective Lagrangian method to calculate their decay widths for the proposed molecular configurations
and compare the results with
experimental data to assess the viability of the molecular interpretation.  Specifically, we investigate whether the $D_{s3}(2860)$ can be interpreted
as a $P$-wave $K^{*}D^{*}$ molecular state~\cite{Hao:2022vwt,Jiang:2026rpu}, and examine the two proposed molecular interpretations of the $D_{sJ}(2860)$, namely, a $P$-wave
$D^{*}K^{*}$ hadronic molecule~\cite{Hao:2022vwt,Jiang:2026rpu} and an $S$-wave $D_1(2420)K$ hadronic molecule~\cite{Guo:2017jvc,Wang:2025jcq,Guo:2011dd}, to determine which
scenario is more likely.

This paper is organized as follows. In Sec.~\ref{Sec: formulism},  we will present the theoretical  formalism.  In Sec.~\ref{Sec: results}, the numerical result will be given,
followed by discussions and conclusions in the last section.

\section{FORMALISM AND INGREDIENTS}\label{Sec: formulism}
\begin{figure}[http]
\begin{center}
\includegraphics[bb=60 150 1050 740, clip, scale=0.45]{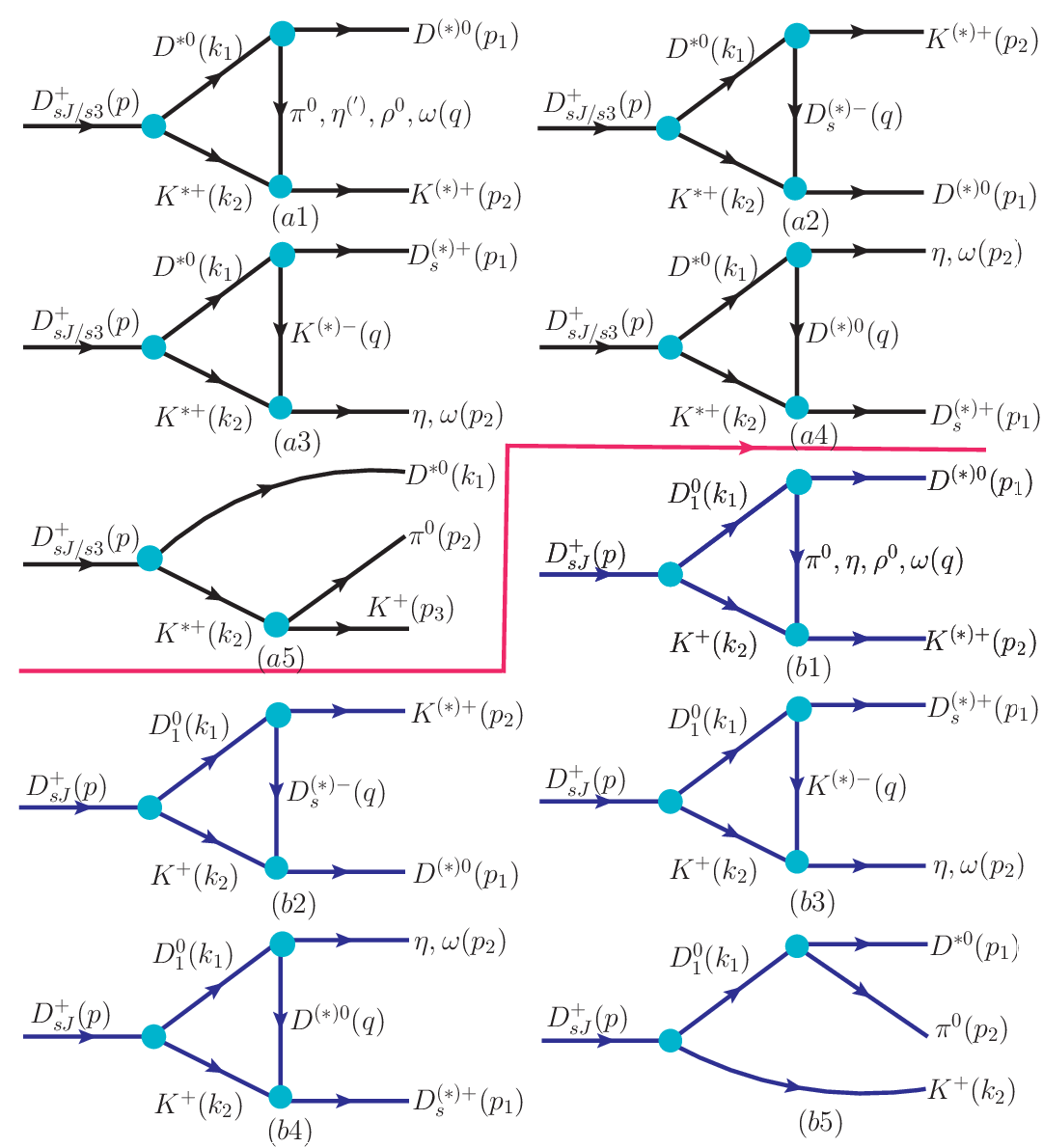}
\caption{Feynman diagrams for the decays of the $D_{sJ}(2860)$ and $D_{s3}(2860)$ within the hadronic molecular framework. The upper panels correspond to the decay diagrams
of these two states assuming the $K^{*}D^{*}$ molecular configuration, while the lower panels show the decay diagrams of the $D_{sJ}(2860)$ as a $D_1K$ molecular state. The
four-momenta of the involved particles are denoted by $(p,k_1,k_2,p_1,p_2,q)$.}
\label{cc1}
\end{center}
\end{figure}

In this work, we investigate the strong decays of the \(D_{sJ}(2860)\) and \(D_{s3}(2860)\), hereafter denoted as \(D_{sJ}\) and \(D_{s3}\), respectively, within the hadronic
molecular framework to examine the viability of their hadronic molecular interpretations.  The corresponding Feynman diagrams are shown in Fig.~\ref{cc1}, where, for simplicity,
only the decay processes of the positively charged states are presented. For the \(D^{*}K^{*}\) molecular configuration, the decay processes of the \(D_{sJ}(2860)\) and \(D_{s3}(2860)\)
into the experimentally observed \(D^{*}K\) and \(DK\) final states are shown in Fig.~\ref{cc1}(a1) and (a2), respectively. The \(D^{*}K\) final state receives contributions
from \(\pi\), \(\eta^{(')}\), \(\rho\), \(\omega\), and \(D_s^{(*)}\) exchanges, while the \(DK\) final state involves \(\rho\), \(\omega\), and \(D_s^{(*)}\) exchanges.
For the \(D_{sJ}(2860)\) interpreted as an \(S\)-wave \(D_1(2420)K\) hadronic molecule, the exchanged particles contributing to the \(D^{*}K\) and \(DK\) decay channels are
almost identical to those in the \(D^{*}K^{*}\) molecular configuration, with the only difference being the absence of the \(\eta'\)-exchange contribution in the \(D^{*}K\)
channel.

In addition, we investigate several decay modes that have not yet been observed experimentally but are expected within the \(D^{(*)}K^{*}\) molecular interpretation of the
\(D_{sJ}(2860)\) and \(D_{s3}(2860)\). These include the \(\eta D_s^{(*)}\) and \(\omega D_s^{(*)}\) decay channels via \(K^{(*)-}\) and \(D^{(*)0}\) exchanges, as well as
the \(DK^{*}\) decay channel via \(\pi\), \(\eta^{(')}\), \(\rho\), \(\omega\), and \(D_s^{(*)}\) exchanges [see Fig.~\ref{cc1}(a3)--(a4)]. Since the width of the \(D^{*}\)
meson is much smaller than that of the \(K^{*}\) meson, we treat the \(D^{*}\) as a stable particle throughout our calculations and also consider the three-body \(K\pi D^{*}\)
decay mode [see Fig.~\ref{cc1}(a5)]. The same decay modes are considered for the \(D_{sJ}(2860)\) under the \(D_1K\) molecular interpretation [see Fig.~\ref{cc1}(b3)--(b5)].

To calculate the Feynman diagrams shown in Fig.~\ref{cc1}, it is necessary to determine the effective Lagrangian densities associated with the relevant interaction vertices.
For the $P$-wave $D_{sJ}D^{*}K^{*}$ and $D_{s3}D^{*}K^{*}$ coupling vertices, we adopt the effective Lagrangian densities used in Refs.~\cite{Huang:2020ptc,Jafarzade:2021vhh},
respectively, which are given by
\begin{align}
{\cal{L}}_{D_{sJ}}&=g_{D_{sJ}}D_{sJ}^{\nu}D^{*}_{\mu}\overleftrightarrow{\partial}_{\nu}K^{*\mu},\label{eq1}\\
{\cal{L}}_{D_{s3}}&=g_{D_{s3}}D_{s3}^{\mu\nu\alpha}D^{*}_{\nu}\overleftrightarrow{\partial}_{\mu}K^{*}_{\alpha}.\label{eq2}
\end{align}
For the $S$-wave $D_{sJ}D_1K$ coupling vertex, the corresponding effective Lagrangian density can be written as~\cite{Yang:2022rck}
\begin{align}
{\cal{L}}_{D_{sJ}}&=g_{D_{sJ}}D_{sJ}^{\mu}D_{1\mu}K,\label{eq3}
\end{align}
where $A\overleftrightarrow{\partial}_{\nu}B=A\partial_{\nu}B-\partial_{\nu}AB$. The coupling constants \(g_{D_{sJ}}\) and \(g_{D_{s3}}\), which characterize the couplings of
the \(D_{sJ}\) and \(D_{s3}\) states to their respective molecular components, will be determined by fitting the experimental data for the \(B_s^0\to\bar{D}^0K^-\pi^+\) decay,
in which the \(D_{sJ}\) and \(D_{s3}\) states were discovered~\cite{LHCb:2014ott}.  Here, we fit only the experimental angular distribution in $\cos\theta(\bar{D}^0K^-)$, with
$\theta$ defined as the angle between the $\pi^+$ and $\bar{D}^0$ momenta in the $\bar{D}^0K^-$ rest frame, in the mass region around $m(\bar{D}^0K^-)\simeq 2860~\mathrm{MeV}/c^2$.
This choice is motivated by the fact that the experimental data in this invariant-mass region are best described by the combined contributions of the spin-1 \(D_{sJ}\) and spin-3
\(D_{s3}\) states~\cite{LHCb:2014ott}.

\begin{figure}[http]
\begin{center}
\includegraphics[bb=70 590 1050 710, clip, scale=0.50]{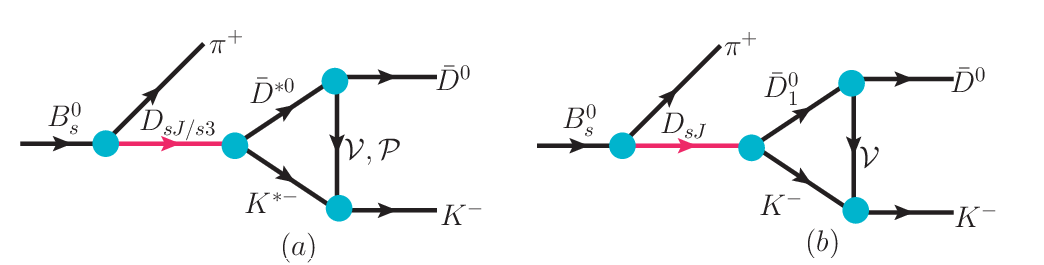}
\caption{Feynman diagrams for the $B_s^0\to\bar{D}^0K^-\pi^+$ decay. Panel (a) corresponds to the scenario in which both $D_{sJ}(2800)$ and $D_{s3}(2800)$ are treated as $D^{*}K^{*}$ molecular states, whereas panel (b) corresponds to the scenario in which $D_{sJ}(2800)$ is treated as a $D_1K$ molecular state. The exchanged vector ${\cal{V}}$ and pseudoscalar ${\cal{P}}$ mesons are specified in Fig.~\ref{cc1}. Note that the final-state $\bar{D}^0$ and $K^-$ can be interchanged.}
\label{cc2}
\end{center}
\end{figure}

The corresponding Feynman diagrams describing the decay $B_s^0\to\bar{D}^0K^-\pi^+$ are shown in Fig.~\ref{cc2}.
For the initial $B_s^0\to\pi^-D_{sJ/s3}^+$ vertices, we employ the effective Lagrangians~\cite{Guo:2022zbc,Jafarzade:2021vhh}
\begin{align}
\mathcal{L}_{B_s^0 D_{sJ} \pi}&= i g_{B_s^0 D_{sJ} \pi }D_{sJ}^\mu\left(\partial_\mu B_s^0 \pi- B_s^0\partial_\mu \pi\right)\label{eq4},\\
\mathcal{L}_{B_s^0 D_{s3} \pi}&= g_{B_s^0 D_{s3} \pi } D_{s3,\mu\nu\rho}B_{s}^0(\partial^{\mu}\partial^{\nu}\partial^{\rho}\pi)\label{eq5},
\end{align}
where $g_{B_s^0 D_{sJ} \pi }$ and $g_{B_s^0 D_{s3} \pi }$ denote the corresponding coupling constants, which will be determined by fitting
the experimental data.

In addition to the interaction vertices described above, we also introduce the effective Lagrangian for the interactions between vector-vector and
vector-pseudoscalar mesons~\cite{Bando:1984ej,Bando:1987br,Nagahiro:2008cv}
\begin{align}
&\mathcal{L}_{vpp} = -i g_{vvp} \langle v_\mu [ p, \partial^\mu p ]\rangle\label{eq8},\\
&\mathcal{L}_{vvp} = \frac{G}{\sqrt{2}}\epsilon^{\mu\nu\alpha\beta}\langle\partial_\mu v_\nu \partial_\alpha v_\beta p\rangle,\\
&\mathcal{L}_{vvv} = i g_v \langle (v_\mu \overleftrightarrow{\partial}_\nu v^\mu ) v^\nu\rangle ,
\end{align}
where $\langle \cdots \rangle$ denotes the trace over the SU(4) flavor space. The coupling constant $g_v=m_{\rho}/(2f_{\pi})$, and  $G$ is defined as
$G=3g'^2/(4\pi^2f)$, with $g'=-G_Vm_\rho/(\sqrt{2}f^2)$~\cite{Nagahiro:2008cv}, where $G_v=55~\mathrm{MeV}$ and $f=93~\mathrm{MeV}$.  $p$ and
$v$ denote the SU(4) pseudoscalar and vector meson fields, respectively. Their explicit forms are given by
\begin{equation}
p =\begin{pmatrix}
\frac{\pi^0}{\sqrt{2}} + \frac{\eta}{\sqrt{3}} + \frac{\eta'}{\sqrt{6}} & \pi^+ & K^+ & \bar{D}^0 \\
\pi^- & -\frac{\pi^0}{\sqrt{2}} + \frac{\eta}{\sqrt{3}} + \frac{\eta'}{\sqrt{6}} & K^0 & D^- \\
K^- & \bar{K}^0 & -\frac{1}{\sqrt{3}}\eta + \frac{2\eta'}{\sqrt{6}} & D_s^- \\
D^0 & D^+ & D_s^+ & \eta_c
\end{pmatrix},
\end{equation}
and
\begin{equation}
v^\mu =
\begin{pmatrix}
\frac{1}{\sqrt{2}}(\rho^0 + \omega) & \rho^+ & K^{*+} & \bar{D}^{*0} \\
\rho^- & \frac{1}{\sqrt{2}}(-\rho^0 + \omega) & K^{*0} & D^{*-} \\
K^{*-} & \bar{K}^{*0} & \phi & D_s^{*-} \\
D^{*0} & D^{*+} & D_s^{*+} & J/\psi
\end{pmatrix}^\mu .
\end{equation}
The coupling $g_{vpp}$ is fixed from the strong decay width of $K^* \to K\pi$.  With the help of Eq.~\ref{eq8}, the two-body decay width $K^{*+} \to K^0 \pi^+$
is related to $g$ as
\begin{equation}
\Gamma(K^{*+} \to K^0 \pi^+) = \frac{g_{vpp}^2}{6\pi m_{K^{*+}}^2}\, P_{\pi K^*}^3 = \frac{2}{3}\,\Gamma_{K^{*+}},
\end{equation}
where $P_{\pi K^*}$ is the three-momentum of the pion in the rest frame of the $K^*$ meson. Using the experimental value of the total decay width $\Gamma_{K^{*+}} = 50.3 \pm 0.8~\mathrm{MeV}$
and the hadron masses~\cite{ParticleDataGroup:2026mpi}, we obtain $g_{vpp} = 4.64$.

Considering the heavy-quark symmetry and chiral symmetry, the relevant effective Lagrangians for $D_1D^{*}{\cal{P}}$ and $D_1D^{(*)}{\cal{V}}$ can be constructed as~\cite{Yue:2024bvy}
\begin{align}
\mathcal{L}_{D_1D^{*}{\cal{P}}}=&g_{D_1D^{*}{\cal{P}}}\Bigg[3D_{1b}^{\mu}\left(\partial_{\mu}\partial_{\nu}{\cal{P}}\right)_{ba}D_{a}^{*\nu\dagger}-D_{1b}^{\mu}
\left(\partial_{\nu}\partial^{\nu}{\cal{P}}\right)_{ba}D_{a\mu}^{*\dagger}\nonumber\\
&+\frac{1}{m_D^{*}m_{D_1}}\left(\partial^{\nu}D_{1b}^{\mu}\right)\left(\partial_{\nu}\partial_{\tau}{\cal{P}}\right)_{ba}\left(\partial^{\tau}D_{a\mu}^{*\dagger}\right)
\Bigg],\\
\mathcal{L}_{D^{(*)} D_1 {\cal{V}}}&=i g_{D^{(*)} D_1 {\cal{V}}}\epsilon_{\mu\nu\alpha\beta}D_{1b}^{\mu}\overleftrightarrow{\partial}^{\alpha}D_{a}^{*\nu\dagger}{\cal{V}}_{ba}^{\beta}\nonumber\\
                                   &+g_{D D_1 {\cal{V}}}D_b D_{1a}^{\mu}{\cal{V}}_{ba}^{\mu},
\end{align}
with $D^{(*)}$= $(D^{(*)0},\,D^{(*)+},\,D_s^{(*)+})$ and $D_1$= $(D_1(2420)^0,\,D_1(2420)^+,\,D_{s1}(2536)^+)$.  The coupling constants are determined by heavy-quark symmetry and chiral
symmetry and are given by~\cite{Yue:2024bvy}
\begin{align}
g_{D^{*} D_1 {\cal{P}}}&=-\frac{\sqrt{6}}{3}\frac{h_1+h_2}{\Lambda_\chi f_\pi}\sqrt{m_{D_1}m_D^{*}},\nonumber\\
g_{D D_1 {\cal{V}}}&=\frac{2g_V\xi_1}{\sqrt{3}}\sqrt{m_Dm_{D_1}},\nonumber\\
g_{D^{*} D_1 {\cal{V}}}&=\frac{g_V\xi_1}{\sqrt{3}}.
\end{align}
with $(h_1+h_2)/\Lambda_\chi=0.55~\mathrm{GeV}^{-1}$, $\xi_1=-0.1$, and $g_V=5.9$.
Here, ${\cal{V}}^{\mu}$ and ${\cal{P}}$ denote the SU(3) vector-meson and pseudoscalar-meson matrices, respectively,
\begin{align}
V_{\mu}&=
\left(
  \begin{array}{ccc}
    \frac{1}{\sqrt{2}}(\rho^{0}+\omega) & \rho^{+}                             &  K^{*+}     \\
    \rho^{-}                            & \frac{1}{\sqrt{2}}(-\rho^{0}+\omega) &  K^{*0}     \\
     K^{*-}                             & \bar{K}^{*0}                         &  \phi       \\
  \end{array}
\right)_{\mu},\\
P&=
\left(
  \begin{array}{ccc}
    \frac{1}{\sqrt{2}}\pi^{0}+\frac{1}{\sqrt{6}}\eta       & \pi^{+}                                                   &  K^{+}                         \\
    \pi^{-}                                                & -\frac{1}{\sqrt{2}}\pi^{0}+\frac{1}{\sqrt{6}}\eta         &  K^0                           \\
     K^{-}                                                 & \bar{K}^{0}                                               &  -\frac{2}{\sqrt{6}}\eta       \\
  \end{array}
\right),
\end{align}

To regularize the ultraviolet divergences in the loop diagrams shown in Figs.~\ref{cc1} and \ref{cc2}, we adopt a four-momentum cutoff scheme
(see Appendix~\ref{section-AApen} for details), with the cutoff parameter \(\Lambda\) determined by fitting the experimental data. Combining
all contributions, we obtain the full decay amplitudes. We first present the amplitude for \(D_{sJ}(2860)\) as a \(D^{*}K^{*}\) molecular state:
\begin{align}
{\cal{M}}_{a1}(&D^{*0}K^{+})=\frac{1}{6\sqrt{2}}[3{\cal{M}}^{+}_P(\pi^0)+2{\cal{M}}^{+}_P(\eta)-{\cal{M}}^{+}_P(\eta^{'})\nonumber\\
                            &+3{\cal{M}}^{+}_V(\rho^0)+3{\cal{M}}^{+}_V(\omega)],\label{eq21}\\
{\cal{M}}_{a2}(&D^{*0}K^{+})=\frac{-1}{\sqrt{2}}[{\cal{M}}^{-}_{P}(D_{s}^{-})+{\cal{M}}^{-}_{V}(D_{s}^{*-})]|^{{D^{*}}\leftrightarrow{}{K^{*}}}_{k_1\leftrightarrow{}k_2},\\
{\cal{M}}_{a1}(&D^{0}K^{+})=\frac{1}{12}[6{\cal{M}}^{+}_{T}(\pi^0)+4{\cal{M}}^{+}_{T}(\eta)-2{\cal{M}}^{+}_{T}(\eta^{'})\nonumber\\
               &+3{\cal{M}}_{Y}(\rho^0)+3{\cal{M}}_{Y}(\omega)],\\
{\cal{M}}_{a2}(&D^{0}K^{+})=-[{\cal{M}}^{-}_{T}(D_{s}^{-})+\frac{1}{2}{\cal{M}}_{Y}(D_{s}^{*-})]|^{{D^{*}}\leftrightarrow{}{K^{*}}}_{k_1\leftrightarrow{}k_2},\\
{\cal{M}}_{a1}(&D^{0}K^{*+})=\frac{1}{2\sqrt{6}}[{\cal{M}}^{-}_{S}(\pi^0,\sqrt{3},0)+{\cal{M}}^{-}_{S}(\eta,\sqrt{2},-\sqrt{2})\nonumber\\
               &+{\cal{M}}^{-}_{S}(\eta^{'},1,2)+\sqrt{3}{\cal{M}}^{+}_{G}(\rho^0)+\sqrt{3}{\cal{M}}^{+}_{G}(\omega)],\\
{\cal{M}}_{a2}(&D^{0}K^{*+})=\frac{-1}{\sqrt{2}}[{\cal{M}}^{+}_{S}(D_s^{-},1,0)+{\cal{M}}^{-}_{G}(D_s^{*-})]|^{{D^{*}}\leftrightarrow{}{K^{*}}}_{k_1\leftrightarrow{}k_2},\\
{\cal{M}}_{a3}(&D_{s}^{+}\omega)=\frac{1}{2}{\cal{M}}^{+}_{G}(K^{*-})+\frac{1}{2}{\cal{M}}^{-}_{S}(K^{-},0,1),\\
{\cal{M}}_{a4}(&D_{s}^{+}\omega)=\frac{1}{2}[{\cal{M}}^{-}_{G}(D^{*0})-{\cal{M}}^{+}_{S}(D^0,1,0)]|^{{D^{*}}\leftrightarrow{}{K^{*}}}_{k_1\leftrightarrow{}k_2},\\
{\cal{M}}_{a3}(&D_{s}^{*+}\eta)=\frac{2}{\sqrt{6}}{\cal{M}}^{-}_{G}(K^{*-})|^{{D^{*}}\leftrightarrow{}{K^{*}}}_{k_1\leftrightarrow{}k_2,p_2\rightarrow{}p_1}+{\cal{M}}^{+}_{P}(K^{-}),\\
{\cal{M}}_{a4}(&D_{s}^{*+}\eta)=\frac{1}{\sqrt{6}}[{\cal{M}}^{+}_{G}(D^{*0})|_{p_2\rightarrow{}p_1}-{\cal{M}}^{-}_{P}(D^0)|^{{D^{*}}\leftrightarrow{}{K^{*}}}_{k_1\leftrightarrow{}k_2}],\\
{\cal{M}}_{a3}(&D_{s}^{+}\eta)=\frac{-1}{\sqrt{3}}{\cal{M}}_{Y}(K^{*-})-\frac{1}{\sqrt{3}}{\cal{M}}^{+}_{T}(K^{-}),
\end{align}
\begin{align}
{\cal{M}}_{a4}(&D_{s}^{+}\eta)=\frac{-1}{2\sqrt{3}}[{\cal{M}}_{Y}(D^{*0})+{\cal{M}}^{+}_{T}(D^0)]|^{{D^{*}}\leftrightarrow{}{K^{*}}}_{k_1\leftrightarrow{}k_2},\label{eq22}\\
{\cal{M}}_{a5}(&D^{*0}\pi^0K^{+})=\frac{-igg_{D_{sJ}}}{\sqrt{2}}\epsilon^{*}_{\sigma}(k_1)(k_2-k_1)\cdot\epsilon{}(p)\nonumber\\
                   &\times{}{\cal{P}}^{\eta\sigma}(k^{*+},k_2)(p_3-p_2)_{\eta},
\end{align}
with
\begin{align}
{\cal{M}}^{\pm}_{P}&(\chi)=-igGg_{D_{sJ}}\int\frac{d^4q}{(2\pi)^4}\epsilon_{\mu\nu\alpha\beta}p_1^{\mu}\epsilon^{*}_{\nu}(p_1)\nonumber\\
                   &\times{}k_1^{\alpha}{\cal{P}}^{\beta\eta}(D^{*0},k_1)(k_2-k_1)\cdot\epsilon{}(p)\nonumber\\
                    &\times{}{\cal{P}}^{\eta\sigma}(K^{*+},k_2)(q\pm{}p_2)^{\sigma}\frac{i}{q^2-m^2_{\chi}},\label{eq34}\\
{\cal{M}}^{\pm}_{V}&(\chi)=igGg_{D_{sJ}}\int\frac{d^4q}{(2\pi)^4}[(k_1\pm{}q)_{\alpha}g_{\beta\eta}\nonumber\\
                   &+(p_1\pm{}(-q))_{\eta}g_{\alpha\beta}-(k_1+p_1)_{\beta}g_{\alpha\eta}]\nonumber\\
                   &\times{}\epsilon^{*\alpha}(p_1){\cal{P}}^{\eta\lambda}(D^{*0},k_1)(k_2-k_1)\cdot\epsilon{}(p)\nonumber\\
                   &\times{}{\cal{P}}^{\lambda\sigma}(K^{*+},k_2)\epsilon_{q\tau{}k_2\sigma}{\cal{P}}^{\beta\tau}(\chi,q),\\
{\cal{M}}^{\pm}_{T}&(\chi)=-igg_{V}g_{D_{sJ}}\int\frac{d^4q}{(2\pi)^4}[p_1\pm{}(-q)]^{\beta}\nonumber\\
                   &\times{}{\cal{P}}^{\beta\eta}(D^{*0},k_1)(k_2-k_1)\cdot\epsilon{}(p)\nonumber\\
                   &\times{}{\cal{P}}^{\eta\sigma}(K^{*+},k_2)(q\pm p_2)^{\sigma}\frac{i}{q^2-m^2_{\chi}},\\
{\cal{M}}_{Y}&(\chi)=-iG^2g_{D_{sJ}}\int\frac{d^4q}{(2\pi)^4}\epsilon_{\mu\eta\alpha\beta}k_1^{\mu}q^{\alpha}{\cal{P}}^{\eta\lambda}(D^{*0},k_1)\nonumber\\
                  &\times{}(k_2-k_1)\cdot\epsilon{}(p){\cal{P}}^{\lambda\sigma}(K^{*+},k_2)\epsilon_{q\tau{}k_2\sigma}{\cal{P}}^{\beta\tau}(\chi,q),\\
{\cal{M}}^{\mp}_{S}&(\chi,a,b)=igGg_{D_{sJ}}\int\frac{d^4q}{(2\pi)^4}(p_1\mp{}q)^{\beta}{\cal{P}}^{\beta\eta}(D^{*0},k_1)\nonumber\\
                   &\times{}(k_2-k_1)\cdot\epsilon{}(p)\frac{i}{q^2-m^2_{\chi}}{\cal{P}}^{\eta\sigma}(K^{*+},k_2)\nonumber\\
                   &\times{}(a\epsilon_{k_2\sigma{}p_2\alpha}+b\epsilon_{p_2\alpha{}k_2\sigma{}})\epsilon^{*}_{\alpha}(p_2),\\
{\cal{M}}^{\pm}_{G}&(\chi)=-igGg_{D_{sJ}}\int\frac{d^4q}{(2\pi)^4}\{[k_2\pm (-q)]_{\alpha}g_{\beta\eta}\nonumber\\
                   &+(p_2\pm q)_{\eta}g_{\alpha\beta}-(k_2+p_2)_{\beta}g_{\alpha\eta}\}\nonumber\\
                   &\times{}\epsilon^{*\alpha}(p_2){\cal{P}}^{\eta\lambda}(k^{*+},k_2)(k_2-k_1)\cdot\epsilon{}(p)\nonumber\\
                   &\times{}{\cal{P}}^{\lambda\sigma}(D^{*0},k_1)\epsilon_{k_1\sigma{}q\tau{}}{\cal{P}}^{\beta\tau}(\chi,q).
\end{align}
where ${\cal{P}}^{\mu\nu}(\chi,p)=i(-g^{\mu\nu}+p^{\mu}p^{\nu}/m^2_{\chi})/(p^2-m^2_{\chi})$.  For the $D_{s3}(2860)$ state in the $D^*K^*$ molecular configuration, the corresponding decay
amplitude has essentially the same form as that for $D_{sJ}(2860)$, with the polarization-vector structure replaced by the polarization tensor of the spin-3 state. Specifically, the factor
$(k_2-k_1)\cdot\epsilon(p)$ is replaced by $(k_2-k_1)\cdot\epsilon^{\eta\tau}(p)$. Further details are provided in Appendix~\ref{sec:aa}.

Next, we present the decay amplitude of the $D_{sJ}(2860)$ state in the $D_1K$ molecular configuration, which is given in detail as follows:
\begin{align}
{\cal{M}}_{b1}(&D^{*0}K^{+})=\frac{-gg_Vg_{D_{sJ}}}{2}\int\frac{d^4q}{(2\pi)^4}\epsilon^{\mu\nu\alpha\beta}\nonumber\\
              &\times{}(p_1+k_1)_{\alpha}\epsilon^{*}_{\nu}(p_1){\cal{P}}^{\mu\lambda}(D_1^0,k_1)\epsilon^{\lambda}(p)\nonumber\\
              &\times{}\frac{i}{k_2^2-m^2_{K^{+}}}(p_2+k_2)_{\tau}{\cal{P}}^{\beta\tau}(\rho^0/\omega,q),\\
{\cal{M}}_{b2}(&D^{*0}K^{+})=\frac{-Gg_{D_1D^{*}P}g_{D_{sJ}}}{2}\int\frac{d^4q}{(2\pi)^4}\epsilon^{\mu\nu\alpha\beta}q_{\mu}p_{1\alpha}\nonumber\\
              &\times{}\epsilon^{*}_{\beta}(p_1)\frac{i}{k_2^2-m^2_{K^{+}}}\epsilon^{\lambda}(p){\cal{P}}^{\lambda\eta}(D_1^0,k_1)(3p_2^{\eta}p_2^{\delta}\nonumber\\
              &-g^{\eta\delta}p_2^2+\frac{g^{\eta\delta}k_1\cdot{}p_2q\cdot{}p_2}{m_{D_1}m_{D^{*}}}){\cal{P}}^{\nu\delta}(D_s^{*-},q),\\
{\cal{M}}_{b1}(&D^{0}K^{+})=\frac{gg_{D_1DV}g_{D_{sJ}}}{2}\int\frac{d^4q}{(2\pi)^4}{\cal{P}}^{\mu\lambda}(D_1^0,k_1)\nonumber\\
                        &\times{}\epsilon^{\lambda}(p)\frac{i}{k_2^2-m^2_{K^{+}}}(p_2+k_2)_{\tau}{\cal{P}}^{\mu\tau}(\rho^0/\omega,q),\\
{\cal{M}}_{b2}(&D^{0}K^{+})=-gg_{D_1D^{*}P}g_{D_{sJ}}\int\frac{d^4q}{(2\pi)^4}(p_1+k_2)^{\mu}\nonumber\\
                        &\times{}\frac{i}{k_2^2-m^2_{K^{+}}}\epsilon^{\lambda}(p){\cal{P}}^{\lambda\eta}(D_1^0,k_1)(3p_2^{\eta}p_2^{\delta}\nonumber\\
                                      &-g^{\eta\delta}p_2^2+\frac{g^{\eta\delta}k_1\cdot{}p_2q\cdot{}p_2}{m_{D_1}m_{D^{*}}}){\cal{P}}^{\mu\delta}(D_s^{*-},q),\\
{\cal{M}}_{b1}(&D^{0}K^{*+})=\frac{Gg_{D_1DV}g_{D_{sJ}}}{2}\int\frac{d^4q}{(2\pi)^4}\nonumber\\
              &\times{}{\cal{P}}^{\mu\lambda}(D_1^0,k_1)\epsilon^{\lambda}(p)\frac{i}{k_2^2-m^2_{K^{+}}}\epsilon^{\delta\tau\alpha\beta}q_{\delta}p_{2\alpha}\nonumber\\
              &\times{}\epsilon^{*}_{\beta}(p_2){\cal{P}}^{\mu\tau}(\rho^0/\omega,q),\\
{\cal{M}}_{b2}(&D^{0}K^{*+})=gg_{D_1D^{*}V}g_{D_{sJ}}\int\frac{d^4q}{(2\pi)^4}\nonumber\\
              &\times{}(p_1+k_2)^{\mu}\frac{i}{k_2^2-m^2_{K^{+}}}\epsilon^{\lambda}(p){\cal{P}}^{\lambda\eta}(D_1^0,k_1)\nonumber\\
              &\times{}\epsilon_{\eta\sigma\tau\delta}(q+k_1)^{\tau}\epsilon_{\delta}^{*}(p_2){\cal{P}}^{\mu\sigma}(D_s^{*-},q),\\
{\cal{M}}_{b3}(&D_s^{+}\omega)=\frac{Gg_{D_1DV}g_{D_{sJ}}}{2}\int\frac{d^4q}{(2\pi)^4}\nonumber\\
&\times{}{\cal{P}}^{\mu\lambda}(D_1^0,k_1)\epsilon^{\lambda}(p)\frac{i}{k_2^2-m^2_{K^{+}}}\epsilon^{\varpi\beta\alpha\tau}q_{\alpha}p_{2\varpi}\nonumber\\
&\times{}\epsilon^{*}_{\beta}(p_2){\cal{P}}^{\mu\tau}(\rho^0/\omega,q),\\
{\cal{M}}_{b3}(&D_s^{*+}\eta)=\frac{-gg_{D_1D^{*}V}g_{D_{sJ}}}{2}\int\frac{d^4q}{(2\pi)^4}\nonumber\\
              &\times{}\epsilon^{\mu\nu\alpha\beta}(p_1+k_1)_{\alpha}\epsilon^{*}_{\nu}(p_1){\cal{P}}^{\mu\lambda}(D_1^0,k_1)\epsilon^{\lambda}(p)\nonumber\\
              &\times{}\frac{i}{k_2^2-m^2_{K^{+}}}(p_2+k_2)_{\tau}{\cal{P}}^{\beta\tau}(K^{*-},q),\\
{\cal{M}}_{b3}(&D_s^{+}\eta)=\frac{gg_{D_1DV}g_{D_{sJ}}}{\sqrt{3}}\int\frac{d^4q}{(2\pi)^4}{\cal{P}}^{\mu\lambda}(D_1^0,k_1)\nonumber\\
              &\times{}\epsilon^{\lambda}(p)\frac{i}{k_2^2-m^2_{K^{+}}}(p_2+k_2)_{\tau}{\cal{P}}^{\mu\tau}(K^{*-},q),\\
\end{align}
\begin{align}
{\cal{M}}_{b4}(&D_s^{+}\omega)=\frac{-gg_{D_1D^{*}V}g_{D_{sJ}}}{\sqrt{2}}\int\frac{d^4q}{(2\pi)^4}\nonumber\\
              &\times{}(p_1+k_2)^{\mu}\frac{i}{k_2^2-m^2_{K^{+}}}\epsilon^{\lambda}(p){\cal{P}}^{\lambda\eta}(D_1^0,k_1)\nonumber\\
              &\times{}\epsilon_{\eta\sigma\tau\delta}(q+k_1)^{\tau}\epsilon_{\delta}^{*}(p_2){\cal{P}}^{\mu\sigma}(D^{*0},q),\\
{\cal{M}}_{b4}(&D_s^{*+}\eta)=\frac{-Gg_{D_1D^{*}P}g_{D_{sJ}}}{2\sqrt{3}}\int\frac{d^4q}{(2\pi)^4}\epsilon_{\mu\nu\alpha\beta}p_1^{\mu}\nonumber\\
              &\times{}\epsilon^{*}_{\nu}(p_1)q^{\alpha}\frac{i}{k_2^2-m^2_{K^{+}}}\epsilon^{\lambda}(p){\cal{P}}^{\lambda\eta}(D_1^0,k_1)(3p_2^{\eta}p_2^{\delta}\nonumber\\
              &-g^{\eta\delta}p_2^2+\frac{g^{\eta\delta}k_1\cdot{}p_2q\cdot{}p_2}{m_{D_1}m_{D^{*}}}){\cal{P}}^{\beta\delta}(D_s^{*-},q),\\
{\cal{M}}_{b4}(&D_s^{+}\eta)=\frac{-gg_{D_1D^{*}P}g_{D_{sJ}}}{\sqrt{6}}\int\frac{d^4q}{(2\pi)^4}(p_1+k_2)^{\mu}\nonumber\\
              &\times{}\frac{i}{k_2^2-m^2_{K^{+}}}\epsilon^{\lambda}(p){\cal{P}}^{\lambda\eta}(D_1^0,k_1)(3p_2^{\eta}p_2^{\delta}\nonumber\\
              &-g^{\eta\delta}p_2^2+\frac{g^{\eta\delta}k_1\cdot{}p_2q\cdot{}p_2}{m_{D_1}m_{D^{*}}}){\cal{P}}^{\mu\delta}(D^{*0},q),\\
{\cal{M}}_{b5}(&D^{*0}\pi^0K^{+})=-\frac{g_{D_1D^{*}P}g_{D_{sJ}}}{\sqrt{2}}(3p_2^{\mu}p_2^{\lambda}\nonumber\\
               &-g^{\mu\lambda}p_2^2+\frac{g^{\mu\lambda}k_1\cdot{}p_2p_1\cdot{}p_2}{m_{D_1}m_{D^{*}}})\epsilon^{*}_{\lambda}(p_1)\nonumber\\
               &\times{}{\cal{P}}^{\mu\eta}(D_1^0,k_1)\epsilon_{\eta}(p).
\end{align}

Using the decay amplitudes for the $D_{sJ}^-$ and $D_{s3}^-$ states decaying into the $\bar{D}^0K^-$ final state obtained above, the decay amplitude for $B_s^0\to\bar{D}^0K^-\pi^+$ can
be constructed as
\begin{align}
{\cal{M}}(B_s^0& \to\bar{D}^0K^-\pi^+)= g_{B^0_{s}D_{sJ}\pi}{\cal{M}}^{\mu}(\bar{D}^0K^-)G^{\mu\nu}_{D_{sJ}}(p)(p_{B_s^0}+p_{\pi})_{\nu} \nonumber\\
                                    &-i g_{B^0_{s}D_{s3}\pi}{\cal{M}}^{\mu\nu\rho}(\bar{D}^0K^-)G^{\mu\nu\rho;\alpha\beta\eta}_{D_{s3}}(p)p_{\pi,\alpha}p_{\pi,\beta}p_{\pi,\eta}.
\end{align}
where $p_{B_s^0}$ and $p_{\pi}$ denote the four-momenta of the $B_s^0$ and $\pi$ mesons, respectively, while $G^{\mu\nu}_{D_{sJ}}$ and $G^{\mu\nu\rho;\alpha\beta\eta}_{D_{s3}}$ represent
the propagators of the $D_{sJ}$ and $D_{s3}$ states, respectively:
\begin{align}
G^{\mu\nu}_{D_{sJ}}&=\frac{i(-g^{\mu\nu}+p^{\mu}p^{\nu}/m^2_{D_{sJ}})}{p^{2}-m_{D_{sJ}}^{2}+i m_{D_{sJ}}\Gamma_{D_{sJ}}},\nonumber\\
G^{\mu\nu\rho,\alpha\beta\eta}_{D_{s3}}&=\frac{i{\cal{N}}^{\mu\nu\rho,\alpha\beta\eta}}{p^{2}-m_{D_{s3}}^{2}+i m_{D_{s3}}\Gamma_{D_{s3}}},\nonumber
\end{align}
with
\begin{align}
{\cal{N}}^{\mu\nu\rho,\alpha\beta\gamma}&={}\frac{1}{15}\Big[G_{\mu\nu}\left(G_{\rho\alpha}G_{\beta\gamma}+G_{\rho\beta}G_{\alpha\gamma}+G_{\rho\gamma}G_{\alpha\beta}\right)\nonumber\\
     &+G_{\mu\rho}\left(G_{\nu\alpha}G_{\beta\gamma}+G_{\nu\beta}G_{\alpha\gamma}+G_{\nu\gamma}G_{\alpha\beta}\right)\nonumber\\
     &+G_{\nu\rho}\left(G_{\mu\alpha}G_{\beta\gamma}+G_{\mu\beta}G_{\alpha\gamma}+G_{\mu\gamma}G_{\alpha\beta}\right)\Big]\nonumber\\
     &-\frac{1}{6}\Big[G_{\mu\alpha}\left(G_{\nu\beta}G_{\rho\gamma}+G_{\nu\gamma}G_{\rho\beta}\right)\nonumber\\
     &+G_{\mu\beta}\left(G_{\nu\alpha}G_{\rho\gamma}+G_{\nu\gamma}G_{\rho\alpha}\right)\nonumber\\
     &+G_{\mu\gamma}\left(G_{\nu\alpha}G_{\rho\beta}+G_{\nu\beta}G_{\rho\alpha}\right)\Big],
\end{align}
where $G_{\mu\nu}=g_{\mu\nu}-k_{\mu}k_{\nu}/k^2$ and $\Gamma_{D_{sJ/s3}}$ denotes the total width of the $D_{sJ}$ or $D_{s3}$ state.

Once the amplitudes are calculated, we can obtain the partial decay widths
\begin{equation}
d\Gamma(D_{sJ/s3}\to{}BC)=\frac{1}{2J+1}\frac{1}{32\pi^2}\frac{|\mathbf{P}_{D_{sJ/s3}}|}{m_{D_{sJ/s3}}^2}|\mathcal{M}_{D_{sJ/s3}\to{}BC}|^2 d\Omega ,
\end{equation}
and
\begin{align}
d\Gamma(D_{sJ/s3}\to{} & BDF)=\frac{1}{2J+1}\frac{1}{(2\pi)^5}\frac{1}{16M_{D_{sJ/s3}}^2} \nonumber\\
&\times |\mathcal{M}_{D_{sJ/s3}\to{}BDF}|^2|\mathbf{p}_D^*||\mathbf{p}_B| dm_{DF} d\Omega_{p_D}^{*}d\Omega_{p_B}\label{eq58} ,
\end{align}
where $J$ is the total angular momentum of the $D_{sJ/s3}$, $|\mathbf{P}_{D_{sJ/s3}}|$ is the three-momentum of the decay products in the center-of-mass frame.
The $(\bm{p}_D^*,\Omega_{p_D}^*)$ are the momentum and angle of the particle $D$ in the rest frame of $D$ and $F$, and $\Omega_{p_B}$ is the angle of the $B$
in the rest frame of the decaying particle.  The $m_{DF}$ is the invariant mass for $D$ and $B$  with $m_D+m_F\leq m_{DF}\leq M-m_B$.  For the angular-dependent
differential decay rate of the $B_s^0\to\bar{D}^0K^-\pi^+$ decay, we obtain the following expression:
\begin{align}
\frac{dBr_{B_s^0\to{}\bar{D}^0K^{-}\pi^{+}}}{d\cos\theta_{\bar{D}^0K^{0}}dm_{\bar{D}^0K^{0}}}&=\frac{1}{128\pi^3m_{B^0_s}^3\Gamma_{B_s^0}}|\mathbf{p}_{\bar{D}^0K^{-}}||\mathbf{p}_{\pi^{+}}|\nonumber\\
&\times{}|\mathcal{M}(m_{\bar{D}^0K^-},\cos\theta_{ \bar{D}^0K^-})|^2 \label{eq59} ,
\end{align}
where $dBr_{B_s^0\to{}\bar{D}^0K^{-}\pi^{+}}=d\Gamma_{B_s^0\to{}\bar{D}^0K^{-}\pi^{+}}/\Gamma_{B_s^0}$~\cite{Guo:2016bjq} with the total decay width of the $B_s^0$ meson is taken as $\Gamma_{B_s^0}=(4.344\pm0.016)\times10^{-4}$~eV~\cite{ParticleDataGroup:2026mpi}, and
\begin{align}
|\mathbf{p}_{\bar{D}^0K^{-}}|=\frac{\lambda^{1/2}(m^2_{\bar{D}^0K^{-}},m^2_{\bar{D}},m^2_{k^{-}})}{2m_{\bar{D}^0K^{-}}};~|\mathbf{p}_{\pi^{+}}|=\frac{\lambda^{1/2}(m^2_{B_{s}^0},
m^2_{\bar{D}^0K^{-}},m^2_{\pi^{+}})}{2m_{\bar{D}^0K^{-}}}.\nonumber
\end{align}
Here, $m_{\bar{D}^0K^-}$ denotes the invariant mass of the $\bar{D}^0K^-$ system and the K\"{a}ll\'{e}n function is defined as
$\lambda(\alpha,\beta,\gamma)=\alpha^2+\beta^2+\gamma^2-2\alpha\beta-2\alpha\gamma-2\beta\gamma$.  Note that Eq.~\ref{eq59} is
derived in the $\bar{D}^0K^-$ rest frame, where the $\pi^+$ meson moves along the positive $z$ axis and the $\bar{D}^0$ meson
moves in the $x$-$z$ plane, with the angle between the momenta of the $\bar{D}^0$ and $\pi^+$ mesons defined as $\theta_{\bar{D}^0K^-}$.

\section{RESULTS AND DISCUSSIONS}\label{Sec: results}
\begin{figure}[http]
\begin{center}
\includegraphics[bb=-80 10 1050 400, clip, scale=0.35]{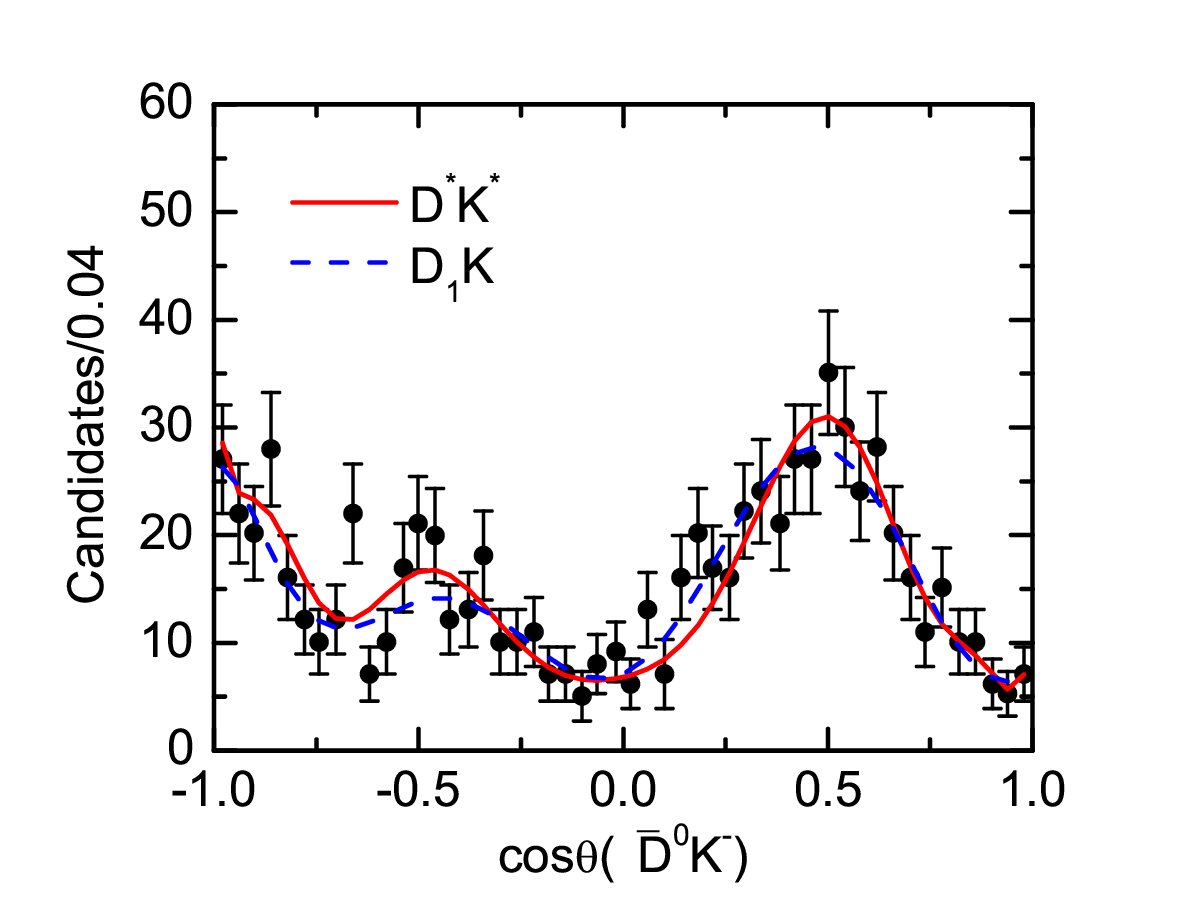}
\caption{The angular distribution of the differential decay rate for the $B_s^0\to\bar{D}^0K^-\pi^+$ decay at $m_{\bar{D}^0K^-}=2860$ MeV. The red solid and blue dashed curves
correspond to the $D^{*}K^{*}$ and $D_1K$ molecular configurations of the $D_{sJ}(2860)$, respectively, while the black dots denote the experimental data from Ref.~\cite{LHCb:2014ott}.}
\label{cc4}
\end{center}
\end{figure}

Within our framework, the four-momentum cutoff \(\Lambda\), the molecular coupling constants \(g_{D_{sJ}}\) and \(g_{D_{s3}}\), and the production coupling
constants \(g_{B_s^0D_{sJ}\pi}\) and \(g_{B_s^0D_{s3}\pi}\), defined in Eqs.~\ref{eq4} and \ref{eq5}, are determined by fitting experimental data. We fit the
measured \(\cos\theta(\bar{D}^0K^-)\) distribution in the \(D_{sJ}(2860)\) and \(D_{s3}(2860)\) mass region of the \(B_s^0\to\bar{D}^0K^-\pi^+\) decay~\cite{LHCb:2014ott},
where \(\theta\) is the angle between the \(\pi^+\) and \(\bar{D}^0\) momenta in the \(\bar{D}^0K^-\) rest frame. Since \(D_{sJ}(2860)\) can be interpreted
as either a \(P\)-wave \(D^*K^*\) molecule or an \(S\)-wave \(D_1K\) molecule, we perform separate fits for the two configurations. We further constrain the
parameters using the experimentally measured partial decay widths associated with the respective \(D_{sJ}(2860)\) and \(D_{s3}(2860)\) contributions~\cite{ParticleDataGroup:2026mpi}.
The results are shown in Fig.~\ref{cc4}: the red solid and blue dashed curves represent the \(D^*K^*\) and \(D_1K\) molecular configurations of \(D_{sJ}(2860)\), respectively,
and the black dots denote the experimental data from Ref.~\cite{LHCb:2014ott}.

As shown in Fig.~\ref{cc4}, both the $P$-wave $D^{*}K^{*}$ and $S$-wave $D_1K$ molecular configurations provide good descriptions of the experimental data. The corresponding reduced
chi-square values, $\chi^2/\mathrm{d.o.f.}$, are $1.75$ and $1.35$, respectively. The fitted parameters for the two configurations are summarized as follows: for the $P$-wave $D^{*}K^{*}$
configuration, $\Lambda^{sJ}=\Lambda^{s3}=1.62$ GeV, $g_{D_{sJ}}=10.80~\mathrm{GeV^{-1}}$, $g_{D_{s3}}=97.13~\mathrm{GeV^{-1}}$, $g_{B_s^0D_{sJ}\pi}=1.65\times{}10^{-8} ~\mathrm{GeV^{-1}}$,
and $g_{B_s^0D_{s3}\pi}=2.44\times{}10^{-9} ~\mathrm{GeV^{-2}}$; while for the $S$-wave $D_1K$ configuration, $\Lambda^{sJ}=\Lambda^{s3}=1.07$ GeV, $g_{D_{sJ}}=17.61~\mathrm{GeV^{-1}}$,
$g_{D_{s3}}=260.06~\mathrm{GeV^{-1}}$, $g_{B_s^0D_{sJ}\pi}=1.15\times{}10^{-7} ~\mathrm{GeV^{-1}}$, and $g_{B_s^0D_{s3}\pi}=9.16\times{}10^{-8} ~\mathrm{GeV^{-2}}$.

\begin{figure}[http]
\begin{center}
\includegraphics[bb=-20 10 1050 280, clip, scale=0.40]{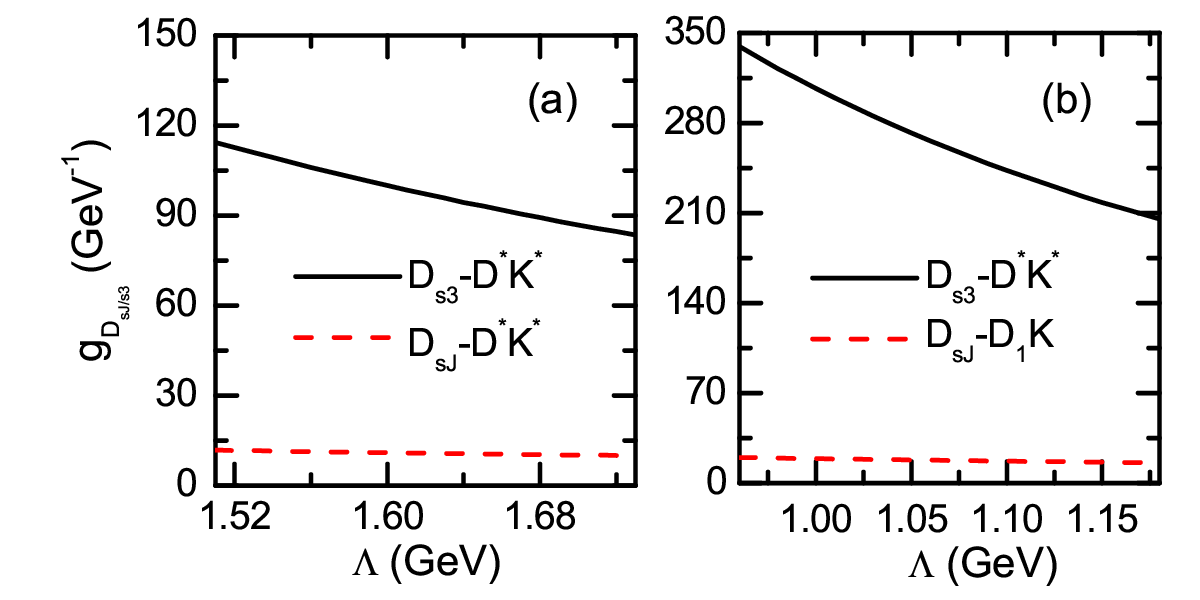}
\caption{The coupling constants as functions of the cutoff parameter $\Lambda$.}
\label{cc5}
\end{center}
\end{figure}

The results presented  above are obtained by fitting the model using the experimental central values of the masses and widthes of $D_{sJ}(2860)$ and $D_{s3}(2860)$, together with the experimental central
values of their respective contributions to the $B_s^0\to\bar{D}^0K^-\pi^+$ decay width~\cite{ParticleDataGroup:2026mpi}.  To account for the experimental uncertainties, we perform fits by simultaneously taking the upper or
lower limits of the masses and decay widths. The resulting fits are presented in Fig.~\ref{cc5}, where we show only the dependence of the coupling constants \(g_{D_{sJ/s3}}\) on the four-momentum
cutoff parameter. Fig.~\ref{cc5}(a) shows the results obtained by interpreting $D_{sJ}(2860)$ as a $P$-wave $D^{*}K^{*}$ molecular state, with $\Lambda$ ranging from  1.51 to 1.73~GeV,
while Fig.~\ref{cc5}(b) shows those obtained by interpreting it as an $S$-wave $D_1K$ molecular state, where $\Lambda$ varies in the range of 0.96--1.18~GeV.

From Fig.~\ref{cc5}, we can see that all the coupling constants decrease with increasing $\Lambda$, but their rates of variation differ significantly. The coupling constants associated with
the $P$-wave $D^{*}K^{*}$ molecular interpretation of the $D_{s3}(2860)$ exhibit the strongest $\Lambda$ dependence, followed by those of the $S$-wave $D_1K$ molecular interpretation
of the $D_{sJ}(2860)$. In contrast, the coupling constants for the $P$-wave $D^{*}K^{*}$ molecular interpretation of the $D_{sJ}(2860)$ show the mildest dependence on $\Lambda$. Numerically,
for the fit in which $D_{sJ}(2860)$ is interpreted as a $D^{*}K^{*}$ molecular state, the coupling constants $g_{D_{sJ}}$ and $g_{D_{s3}}$ decrease from $11.79~\mathrm{GeV}^{-1}$ and
$114.38~\mathrm{GeV}^{-1}$ to $9.93~\mathrm{GeV}^{-1}$ and $83.46~\mathrm{GeV}^{-1}$, respectively, as $\Lambda$ varies over the considered range.  In contrast, fitting the $D_{sJ}(2860)$ as
a $D_1K$ molecular state yields maximum and minimum values of $339.39~\mathrm{GeV}^{-1}$ and $205.20~\mathrm{GeV}^{-1}$, respectively, for the coupling constant associated with the $P$-wave
$D^{*}K^{*}$ molecular interpretation of $D_{s3}(2860)$, while $g_{D_{sJ}}$ varies from $19.87$ to $15.76~\mathrm{GeV}^{-1}$.

\begin{figure}[http]
\begin{center}
\includegraphics[bb=-80 40 1050 470, clip, scale=0.40]{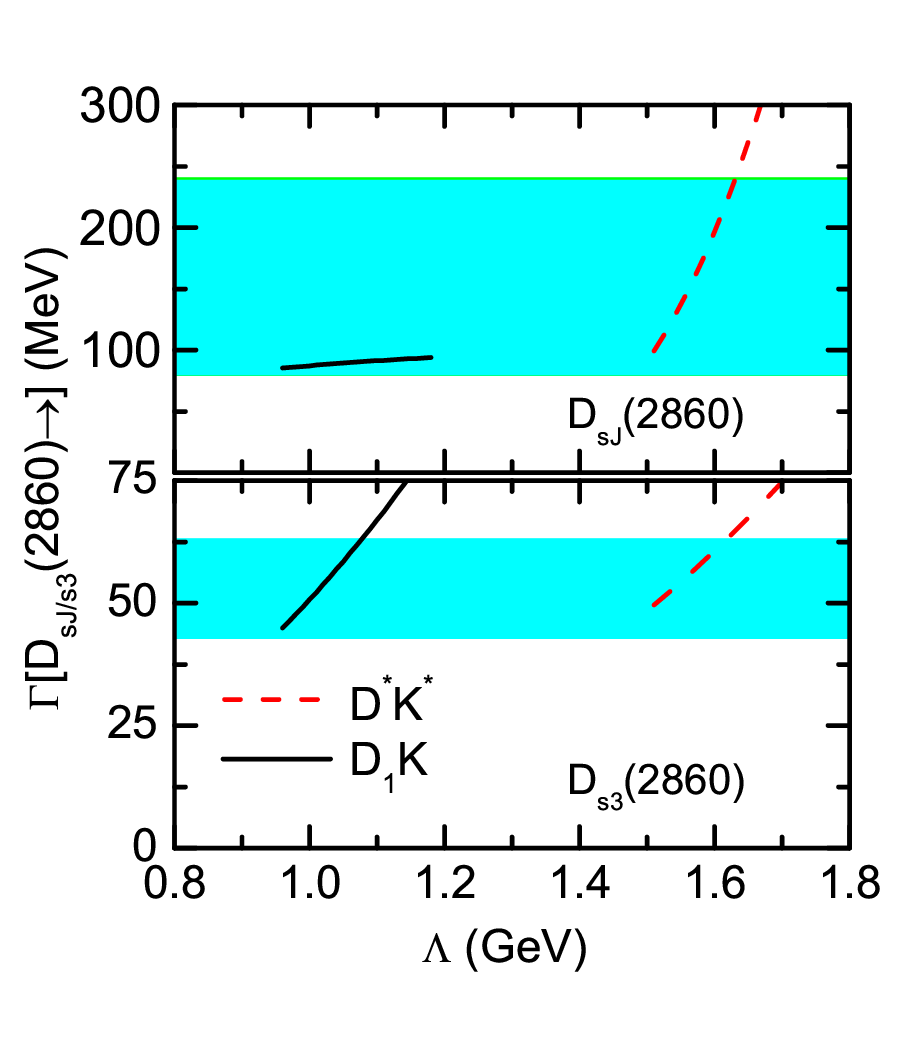}
\caption{Theoretically calculated total decay widths of $D_{s3}(2860)$ (lower panel) and $D_{sJ}(2860)$ (upper panel) as functions of the four-momentum cutoff parameter $\Lambda$. The black solid curves represent the decay widths obtained using the coupling constants and cutoff parameters extracted by treating $D_{sJ}(2860)$ as an $S$-wave $D_1K$ molecular state, while the red dashed curves correspond to the results obtained by treating it as a $P$-wave $D^*K^*$ molecular state. The cyan bands indicate the experimentally measured decay widths of $D_{sJ}(2860)$ and $D_{s3}(2860)$, respectively~\cite{ParticleDataGroup:2026mpi}.}
\label{cc6}
\end{center}
\end{figure}

With the obtained coupling constants and four-momentum cutoff parameters, we calculate the decay widths of $D_{sJ}(2860)$ and $D_{s3}(2860)$ under the molecular interpretations involving $D^{*}K^{*}$
or $D_1K$, respectively. The corresponding results are presented in Fig.~\ref{cc6}, where the cyan band represents the experimental measurements~\cite{ParticleDataGroup:2026mpi}, including their uncertainties.

We can find that, when $D_{sJ}(2860)$ is interpreted as an $S$-wave $D_1K$ molecular state, its calculated total decay width depends only weakly on the cutoff parameter $\Lambda$, increasing
slightly as $\Lambda$ increases. By contrast, the predicted total decay width of $D_{s3}(2860)$ shows a much stronger dependence on $\Lambda$. Specifically, as $\Lambda$ varies from $0.96$
to $1.18$ GeV, the decay width of $D_{sJ}(2860)$ increases from $85.151$ to $93.819$ MeV, corresponding to a variation of $8.668$ MeV. In comparison, the decay width of $D_{s3}(2860)$
increases from $44.849$ to $81.710$ MeV, corresponding to a variation of $36.861$ MeV,  approximately $4.253$ times larger than that of $D_{sJ}(2860)$. This clearly indicates that the
predicted decay width of $D_{s3}(2860)$ is considerably more sensitive to the cutoff parameter than that of $D_{sJ}(2860)$.

Comparing these predictions with the experimental data, we find that the calculated decay width of $D_{sJ}(2860)$ remains in good agreement with the experimental value throughout the
fitted range of $\Lambda=0.96$--$1.18$ GeV. In contrast, the predicted decay width of $D_{s3}(2860)$ increases more rapidly and exceeds the upper limit of its experimental range for
$\Lambda>1.08$ GeV. Nevertheless, the agreement with the experimental data over an appropriate range of $\Lambda$ suggests that the $S$-wave $D_1K$ molecular interpretation of $D_{sJ}(2860)$
remains viable. Moreover, our results are compatible with interpreting $D_{s3}(2860)$ as a $P$-wave $D^*K^*$ molecular state.

When $D_{sJ}(2860)$ is interpreted as a $P$-wave $D^*K^*$ molecular state, the calculated total decay widths of both $D_{sJ}(2860)$ and $D_{s3}(2860)$ exhibit a pronounced dependence
on the cutoff parameter $\Lambda$, increasing rapidly with it. Within the fitted range of $\Lambda=1.51$--$1.73$ GeV, the predicted widths of both states exceed their corresponding
experimental upper limits over certain intervals. In particular, the calculated decay width of $D_{sJ}(2860)$ reaches $240.341$ MeV at $\Lambda=1.636$ GeV, slightly exceeding its
experimental upper limit of $240$ MeV. For $D_{s3}(2860)$, the predicted decay width exceeds the experimental value of $63$ MeV for $\Lambda>1.620$ GeV; at $\Lambda=1.620$ GeV, it
is already $63.240$ MeV. Despite these deviations in parts of the fitted parameter range, the theoretical predictions remain consistent with the experimental data over other portions
of this range. Therefore, the present results still provide support for interpreting both $D_{sJ}(2860)$ and $D_{s3}(2860)$ as $P$-wave $D^*K^*$ molecular states.

By comparing the calculated total decay widths under the two molecular assignments of $D_{sJ}(2860)$, namely, the $S$-wave $D_1K$ and $P$-wave $D^*K^*$ configurations, with the
experimental data, we find that the $S$-wave $D_1K$ molecular interpretation provides a slightly better description of the data. This is because the calculated total decay widths
remain within the experimental range throughout the entire range of $\Lambda$ obtained from the fit under the $S$-wave $D_1K$ molecular assumption. This suggests that the present
analysis based on the total decay widths favors interpreting $D_{sJ}(2860)$ as an $S$-wave $D_1K$ molecular state.

\begin{figure}[http]
\begin{center}
\includegraphics[bb=-30 10 1050 400, clip, scale=0.40]{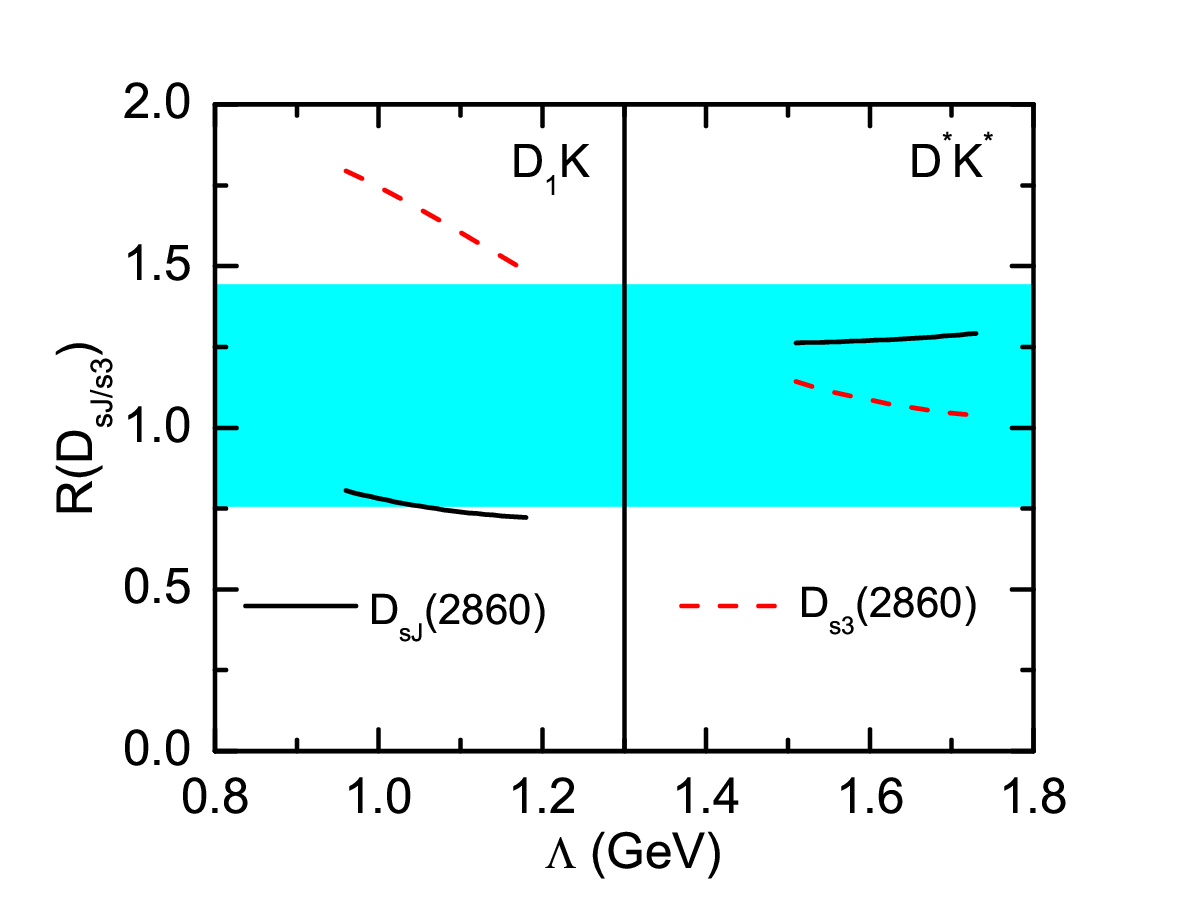}
\caption{Theoretically calculated ratios $R$ (see Eq.~\ref{eq1-1}) of the partial decay widths of $D_{sJ}(2860)$ (black solid curves) and $D_{s3}(2860)$ (red dashed curves) into $KD^*$ and $KD$ as functions of the cutoff parameter $\Lambda$. The left and right panels correspond to the cases where $D_{sJ}(2860)$ is interpreted as an $S$-wave $D_1K$ molecule and a $P$-wave $D^*K^*$ molecule, respectively. Note that $D_{s3}(2860)$ is consistently treated as a $P$-wave $D^*K^*$ molecular state in both cases. The cyan band represents the experimentally measured ratio of the partial decay widths of $D_{s3}(2860)$ into $KD^*$ and $KD$~\cite{ParticleDataGroup:2026mpi}.}
\label{cc7}
\end{center}
\end{figure}
However, our calculations of the ratio $R$ of the partial decay widths into the $KD^*$ and $KD$ channels, considering two scenarios in which $D_{sJ}(2860)$ and $D_{s3}(2860)$ are
interpreted as $D_1K$ and $D^*K^*$ molecular states, respectively, or both states are interpreted as $D^*K^*$ molecular states, reveal that only the latter scenario yields a prediction
for $R_{D_{s3}(2860)}$ that can be compared with the experimental measurement~\cite{ParticleDataGroup:2026mpi}.  The detailed numerical results and their comparison with the experimental
data are presented in Fig.~\ref{cc7}, where the black solid curves and red dashed curves represent $R_{D_{sJ}(2860)}$ and $R_{D_{s3}(2860)}$, respectively.  The cyan band represents the
experimentally measured ratio of the partial decay widths of $D_{s3}(2860)$ into $KD^*$ and $KD$~\cite{ParticleDataGroup:2026mpi}.

As shown in Fig.~\ref{cc7}, when $D_{sJ}(2860)$ and $D_{s3}(2860)$ are interpreted as $S$-wave $D_1K$ and $P$-wave $D^*K^*$ molecular states, respectively, the predicted ratio $R_{D_{s3}(2860)}$ decreases from $1.795$ to $1.487$ as $\Lambda$ varies from $0.96$ to $1.18$ GeV. However, all predicted values lie above the upper limit of the experimental range, disfavoring this combined molecular assignment. In contrast, when both states are assigned as $P$-wave $D^*K^*$ molecules, the predicted ratio decreases from $1.143$ to $1.037$ as $\Lambda$ increases from $1.51$ to $1.73$ GeV, remaining in good agreement with the experimental range throughout this interval. Meanwhile, the predicted ratio $R_{D_{sJ}(2860)}$, which has not yet been measured experimentally, increases gradually from $1.263$ to $1.293$ over the same cutoff range.

Overall, our results indicate that the total decay widths and partial-width ratios provide complementary constraints on the molecular interpretations of $D_{sJ}(2860)$ and $D_{s3}(2860)$. While the calculated total decay widths slightly favor the $S$-wave $D_1K$ assignment for $D_{sJ}(2860)$, the corresponding prediction for $R_{D_{s3}(2860)}$ is in significant disagreement with experiment. Conversely, the $P$-wave $D^*K^*$ assignment for both states better reproduces the measured ratio, although its description of the total decay widths is less satisfactory. This tension indicates that neither observable alone is sufficient to determine the internal structures of these states. A combined analysis of the total decay widths and partial-width ratios, together with more precise future measurements, is therefore essential to further test the proposed molecular assignments. In particular, we strongly encourage future measurements of $R_{D_{sJ}(2860)}$, which would provide an important additional test of the $P$-wave $D^*K^*$ molecular interpretation of both states.

\section{Summary}\label{sec:summary}
In this work, we investigate the strong decay properties of $D_{sJ}(2860)$ and $D_{s3}(2860)$ within the hadronic molecular framework using the effective Lagrangian approach. Specifically, we consider two possible molecular configurations for $D_{sJ}(2860)$, namely, an $S$-wave $D_1K$ configuration and a $P$-wave $D^*K^*$ configuration, while treating $D_{s3}(2860)$ as a $P$-wave $D^*K^*$ molecule. By fitting the angular distributions of the $B_s^0\to\bar{D}^0K^-\pi^+$ decay and the decay widths contributed by these two states, we determine the corresponding coupling constants and four-momentum cutoff parameter $\Lambda$. With these parameters, we calculate their total decay widths and the ratios of their partial decay widths into the $D^*K$ and $DK$ channels.

We first compare the calculated total decay widths with the experimental data. The results indicate that the $S$-wave $D_1K$ molecular interpretation of $D_{sJ}(2860)$ provides a slightly better description of the measured widths than the $P$-wave $D^*K^*$ assignment. In the former scenario, the predicted total width of $D_{sJ}(2860)$ remains within the experimental range throughout the fitted interval of $\Lambda=0.96$--$1.18$ GeV. Although the width of $D_{s3}(2860)$ exhibits a stronger dependence on $\Lambda$ and exceeds the experimental upper limit for $\Lambda>1.08$ GeV, it remains compatible with the data over part of the fitted interval. In the latter scenario, the predicted total widths of both states show a more pronounced dependence on $\Lambda$ and exceed the corresponding experimental upper limits in certain portions of the fitted interval, while remaining consistent with the data elsewhere. Overall, the total decay widths slightly favor the $S$-wave $D_1K$ interpretation of $D_{sJ}(2860)$, although the $P$-wave $D^*K^*$ assignment cannot be ruled out completely.

We further compare the ratio $R_{D_{s3}(2860)}$ of the partial decay widths of $D_{s3}(2860)$ into the $KD^*$ and $KD$ channels with the experimental data.  Interestingly, when $D_{sJ}(2860)$ and $D_{s3}(2860)$ are interpreted as $D_1K$ and $D^*K^*$ molecular states, respectively, all predicted values of $R_{D_{s3}(2860)}$ lie above the upper limit of the experimental range. In contrast, when both states are assigned as $P$-wave $D^*K^*$ molecules, our predictions for $R_{D_{s3}(2860)}$ remain in good agreement with the experimental measurement throughout the fitted interval of $\Lambda=1.51$--$1.73$ GeV. In this scenario, the predicted ratio $R_{D_{sJ}(2860)}$ varies from $1.263$ to $1.293$ over the same interval.  Future measurements of $R_{D_{sJ}(2860)}$ will provide important tests of the proposed molecular configurations. These results indicate that, although the total decay widths favor the $S$-wave $D_1K$ interpretation of $D_{sJ}(2860)$, the partial-width ratios provide complementary constraints and may help clarify whether both $D_{sJ}(2860)$ and $D_{s3}(2860)$ can be consistently described as $P$-wave $D^*K^*$ molecular states.

\section*{Acknowledgments}
This work was supported by the Sailing Plan Project of Yibin University (No. 2021QH06). Y. Huang acknowledges support from the National Natural Science Foundation of China under Grant No. 12005177,
as well as support from the Fundamental Research Funds for the Central Universities under Grant No. 2682026TPY011.
2682026TPY011.

\section{Appendix-A}\label{sec:aa}
\begin{align}
{\cal{M}}^{\pm}_P&(\chi)=-igGg_{D_{s3}}\int\frac{d^4q}{(2\pi)^4}\epsilon_{\mu\nu\alpha\beta}p_1^{\mu}\epsilon^{*}_{\nu}(p_1)\nonumber\\
                   &\times{}k_1^{\alpha}\frac{i(-g^{\beta\eta}+k_1^{\beta}k_1^{\eta}/m^2_{D^{*0}})}{k_1^2-m^2_{D^{*0}}}(k_2-k_1)\cdot\epsilon^{\eta\tau}(p)\nonumber\\
                   &\times{}\frac{i(-g^{\tau\sigma}+k_2^{\tau}k_2^{\sigma}/m^2_{K^{*+}})}{k_2^2-m^2_{K^{*+}}}(q\pm{}p_2)^{\sigma}\frac{i}{q^2-m^2_{\chi}}, \nonumber\\
{\cal{M}}^{\pm}_{V}&(\chi)=igGg_{D_{sJ}}\int\frac{d^4q}{(2\pi)^4}[(k_1\pm{}q)_{\alpha}g_{\beta\eta}\nonumber\\
                   &+(p_1\pm{}(-q))_{\eta}g_{\alpha\beta}-(k_1+p_1)_{\beta}g_{\alpha\eta}]\epsilon^{*\alpha}(p_1)\nonumber\\
                   &\times{}\frac{i(-g^{\eta\lambda}+k_1^{\eta}k_1^{\lambda}/m^2_{D^{*0}})}{k_1^2-m^2_{D^{*0}}}(k_2-k_1)\cdot\epsilon^{\lambda\varphi}(p)\nonumber\\
                   &\times{}\frac{i(-g^{\varphi\sigma}+k_2^{\varphi}k_2^{\sigma}/m^2_{K^{*+}})}{k_2^2-m^2_{K^{*+}}}\epsilon_{q\tau{}k_2\sigma}\frac{i(-g^{\beta\tau}+q^{\beta}q^{\tau}/m^2_{\chi})}{q^2-m^2_{\chi}},\nonumber\\
{\cal{M}}^{\pm}_{T}&(\chi)=-ig^2g_{D_{sJ}}\int\frac{d^4q}{(2\pi)^4}[p_1\pm{}(-q)]^{\beta}\nonumber\\
                   &\times{}\frac{i(-g^{\beta\eta}+k_1^{\beta}k_1^{\eta}/m^2_{D^{*0}})}{k_1^2-m^2_{D^{*0}}}(k_2-k_1)\cdot\epsilon^{\eta\tau}(p)\nonumber\\
                   &\times{}\frac{i(-g^{\tau\sigma}+k_2^{\tau}k_2^{\sigma}/m^2_{K^{*+}})}{k_2^2-m^2_{K^{*+}}}(q\pm{}p_2)^{\sigma}\frac{i}{q^2-m^2_{\chi}},\\
{\cal{M}}_{Y}&(\chi)=-iG^2g_{D_{sJ}}\int\frac{d^4q}{(2\pi)^4}\epsilon_{\mu\eta\alpha\beta}k_1^{\mu}q^{\alpha}\nonumber\\
                   &\times{}\frac{i(-g^{\eta\lambda}+k_1^{\eta}k_1^{\lambda}/m^2_{D^{*0}})}{k_1^2-m^2_{D^{*0}}}(k_2-k_1)\cdot\epsilon^{\eta\varphi}(p)\nonumber\\
                    &\times{}\frac{i(-g^{\varphi\sigma}+k_2^{\varphi}k_2^{\sigma}/m^2_{K^{*+}})}{k_2^2-m^2_{K^{*+}}}\epsilon_{q\tau{}k_2\sigma}\frac{i(-g^{\beta\tau}+q^{\beta}q^{\tau}/m^2_{\chi})}{q^2-m^2_{\chi}},\nonumber  \\
{\cal{M}}^{\mp}_{S}&(\chi,a,b)=igGg_{D_{sJ}}\int\frac{d^4q}{(2\pi)^4}(p_1\mp{}q)^{\beta}\nonumber\\
                   &\times{}\frac{i(-g^{\beta\eta}+k_1^{\beta}k_1^{\eta}/m^2_{D^{*0}})}{k_1^2-m^2_{D^{*0}}}(k_2-k_1)\cdot\epsilon^{\eta\varphi}(p)\frac{i}{q^2-m^2_{\chi}}\nonumber\\
                   &\times{}\frac{i(-g^{\varphi\sigma}+k_2^{\varphi}k_2^{\sigma}/m^2_{K^{*+}})}{k_2^2-m^2_{K^{*+}}}(a\epsilon_{k_2\sigma{}p_2\beta}+b\epsilon_{p_2\beta{}k_2\sigma{}})\epsilon^{*}_{\beta}(p_2),\nonumber
\end{align}
\begin{align}
{\cal{M}}^{\pm}_{G}&(\chi)=-igGg_{D_{sJ}}\int\frac{d^4q}{(2\pi)^4}\{[k_2\pm (-q)]_{\alpha}g_{\beta\eta}\nonumber\\
                   &+(p_2\pm q)_{\eta}g_{\alpha\beta}-(k_2+p_2)_{\beta}g_{\alpha\eta}\}\epsilon^{*\alpha}(p_2)\nonumber\\
                   &\times{}\frac{i(-g^{\eta\lambda}+k_2^{\eta}k_2^{\lambda}/m^2_{K^{*+}})}{k_2^2-m^2_{K^{*+}}}(k_2-k_1)\cdot\epsilon^{\lambda\varphi}{}(p)\nonumber\\
                   &\times{}\frac{i(-g^{\varphi\sigma}+k_1^{\varphi}k_1^{\sigma}/m^2_{D^{*0}})}{k_1^2-m^2_{D^{*0}}}\epsilon_{k_1\sigma{}q\tau{}}\frac{i(-g^{\beta\tau}+q^{\beta}q^{\tau}/m^2_{\chi})}{q^2-m^2_{\chi}},\nonumber
\end{align}

\section{Appendix-B}\label{section-AApen}
We first introduce the Feynman parametrization
\begin{align}
\frac{1}{abc}
=
2\int_0^1 dx\int_0^{1-x}dy\,
\frac{1}{\left[xa+yb+(1-x-y)c\right]^3},
\end{align}
where $z=1-x-y$.
For the three-point loop integral, we consider
\begin{align}
A^{\mu\nu\alpha\beta}
&=
\int\frac{d^4q}{(2\pi)^4}
\frac{q^{\mu}q^{\nu}q^{\alpha}q^{\beta}}
{q^2-m_{\pi}^2+i\epsilon}
\frac{1}{(p_1+q)^2-m_{D^{*0}}^2+i\epsilon}
\nonumber\\
&\quad\times
\frac{1}{(p_2-q)^2-m_{K^{*+}}^2+i\epsilon}.
\end{align}
For convenience, we define
\begin{align}
a&=q^2-m_1^2+i\epsilon,\nonumber\\
b&=(p_1+q)^2-m_2^2+i\epsilon,\nonumber\\
c&=(p_2-q)^2-m_3^2+i\epsilon.
\end{align}
The combined denominator then reads
\begin{align}
xa+yb+zc&=q^2+2(yp_1-zp_2)\cdot q+y(p_1^2-m_2^2)\nonumber\\
        &+z(p_2^2-m_3^2)-xm_1^2.
\end{align}
Defining $Q=yp_1-zp_2$, and performing the momentum shift $l=q+Q$, we obtain
\begin{align}
xa+yb+zc&=l^2-\mathcal{M},
\end{align}
where
\begin{align}
\mathcal{M}&=y^2p_1^2+z^2p_2^2-2yz\,p_1\cdot p_2-y(p_1^2-m_2^2)\nonumber\\
&-z(p_2^2-m_3^2)+xm_1^2.
\end{align}
For the corresponding scalar three-point integral, we then have
\begin{align}
A&=2\int_0^1 dx\int_0^{1-x}dy\int\frac{d^4l}{(2\pi)^4}\frac{1}{(l^2-\mathcal{M}+i\epsilon)^3}.
\end{align}

After the Feynman parametrization and the momentum shift \(l=q+Q\), the rank-four tensor integral can be written as
\begin{align}
A^{\mu\nu\alpha\beta}&=2\int_0^1 dx\int_0^{1-x}dy\int\frac{d^4l}{(2\pi)^4}\nonumber\\
                     &\times\frac{(l-Q)^{\mu}(l-Q)^{\nu}(l-Q)^{\alpha}(l-Q)^{\beta}}{(l^2-\mathcal{M}+i\epsilon)^3}.
\end{align}
Expanding the numerator and using the fact that terms containing an odd number of \(l^\mu\) vanish upon symmetric integration, we obtain
\begin{align}
A^{\mu\nu\alpha\beta}&=2\int_0^1 dx\int_0^{1-x}dy\int\frac{d^4l}{(2\pi)^4}\frac{1}{(l^2-\mathcal{M}+i\epsilon)^3}\nonumber\\
&\times\Big[l^\mu l^\nu l^\alpha l^\beta+\left(Q^\mu Q^\nu l^\alpha l^\beta+\text{5 permutations}\right)+Q^\mu Q^\nu Q^\alpha Q^\beta\Big].
\end{align}
Using Lorentz symmetry, the tensor integrals can be reduced to scalar integrals according to
\begin{align}
\int\frac{d^4l}{(2\pi)^4}\frac{l^\alpha l^\beta}{(l^2-\mathcal{M}+i\epsilon)^3}
&=\frac{g^{\alpha\beta}}{4}\int\frac{d^4l}{(2\pi)^4}\frac{l^2}{(l^2-\mathcal{M}+i\epsilon)^3},
\end{align}
and
\begin{align}
\int\frac{d^4l}{(2\pi)^4}\frac{l^\mu l^\nu l^\alpha l^\beta}{(l^2-\mathcal{M}+i\epsilon)^3}
&=\frac{g^{\mu\nu}g^{\alpha\beta}+g^{\mu\alpha}g^{\nu\beta}+g^{\mu\beta}g^{\nu\alpha}}{24}\nonumber\\
&\times\int\frac{d^4l}{(2\pi)^4}\frac{(l^2)^2}{(l^2-\mathcal{M}+i\epsilon)^3}.
\end{align}
Therefore, the rank-four tensor integral is reduced to a combination of scalar loop integrals.

We next derive the expression for the four-momentum cutoff regularization. As an example, we consider the loop integral involving \(l^4\), namely,

$$
I_4(\mathcal{M})
=
\int\frac{d^4l}{(2\pi)^4}
\frac{(l^2)^2}
{(l^2-\mathcal{M}+i\epsilon)^3}.
$$

Performing the Wick rotation,

$$
l^0=i l_E^0,\qquad
d^4l=i\,d^4l_E,\qquad
l^2=-l_E^2,
$$

we obtain

$$
I_4(\mathcal{M})
=
-i\int\frac{d^4l_E}{(2\pi)^4}
\frac{l_E^4}{(l_E^2+\mathcal{M})^3}.
$$

We then introduce a four-momentum cutoff,

$$
l_E^2\leq\Lambda^2.
$$

Using the four-dimensional spherical measure,

$$
d^4l_E=2\pi^2 l_E^3\,dl_E,
$$

the integral becomes

$$
I_4(\mathcal{M})
=
-\frac{i}{8\pi^2}
\int_0^\Lambda dl_E\,
\frac{l_E^7}{(l_E^2+\mathcal{M})^3}.
$$

Defining \(t=l_E^2+\mathcal{M}\), we have

$$
I_4(\mathcal{M})=-\frac{i}{16\pi^2}\int_{\cal{M}}^{{\cal{M}}+\Lambda^2}dt\,\frac{(t-{\cal{M}})^3}{t^3}.
$$

The remaining integral can be evaluated analytically, yielding

\begin{align}
I_4(\mathcal{M})&=-\frac{i}{16\pi^2}[\Lambda^2-3{\cal{M}}\ln{(1+\frac{\Lambda^2}{{\cal{M}}})}-\frac{3{\cal{M}}^2}{\Lambda^2+{\cal{M}}}\nonumber\\
                &+\frac{{\cal{M}}^2}{2({\cal{M}}+\Lambda^2)^2}+\frac{5{\cal{M}}}{2}].
\end{align}

%\end{CJK*}
%

\begin{thebibliography}{23}%
%\cite{ParticleDataGroup:2026mpi}
\bibitem{ParticleDataGroup:2026mpi}
F.~Takahashi \textit{et al.} [Particle Data Group],
%``Review of Particle Physics,''
Int. J. Mod. Phys. A \textbf{41} (2026), 2630011.

%\cite{Godfrey:1985xj}
\bibitem{Godfrey:1985xj}
S.~Godfrey and N.~Isgur,
%``Mesons in a Relativized Quark Model with Chromodynamics,''
Phys. Rev. D \textbf{32}, 189-231 (1985).
%3113 citations counted in INSPIRE as of 29 May 2023


%\cite{Koniuk:1979vy}
\bibitem{Koniuk:1979vy}
R.~Koniuk and N.~Isgur,
%``Baryon Decays in a Quark Model with Chromodynamics,''
Phys. Rev. D \textbf{21}, 1868 (1980)
[erratum: Phys. Rev. D \textbf{23}, 818 (1981)].
%621 citations counted in INSPIRE as of 29 May 2023


%\cite{Isgur:1978wd}
\bibitem{Isgur:1978wd}
N.~Isgur and G.~Karl,
%``Positive Parity Excited Baryons in a Quark Model with Hyperfine Interactions,''
Phys. Rev. D \textbf{19}, 2653 (1979)
[erratum: Phys. Rev. D \textbf{23}, 817 (1981)].
%1065 citations counted in INSPIRE as of 29 May 2023



%\cite{Belle:2003nnu}
\bibitem{Belle:2003nnu}
S.~K.~Choi \textit{et al.} [Belle],
%``Observation of a narrow charmonium-like state in exclusive $B^\pm \to K^\pm \pi^+ \pi^- J/\psi$ decays,''
Phys. Rev. Lett. \textbf{91} (2003), 262001.


%\cite{LHCb:2015yax}10
\bibitem{LHCb:2015yax}
R.~Aaij \textit{et al.} [LHCb],
%``Observation of $J/\psi p$ Resonances Consistent with Pentaquark States in $\Lambda_b^0 \to J/\psi K^- p$ Decays,''
Phys. Rev. Lett. \textbf{115}, 072001 (2015).
%1658 citations counted in INSPIRE as of 28 Dec 2023


%\cite{LHCb:2016ztz}
\bibitem{LHCb:2016ztz}
R.~Aaij \textit{et al.} [LHCb],
%``Model-independent evidence for $J/\psi p$ contributions to $\Lambda_b^0\to J/\psi p K^-$ decays,''
Phys. Rev. Lett. \textbf{117}, 082002 (2016).
%187 citations counted in INSPIRE as of 28 Dec 2023


%\cite{LHCb:2016lve}
\bibitem{LHCb:2016lve}
R.~Aaij \textit{et al.} [LHCb],
%``Evidence for exotic hadron contributions to $\Lambda_b^0 \to J/\psi p \pi^-$ decays,''
Phys. Rev. Lett. \textbf{117},082003 (2016).
%161 citations counted in INSPIRE as of 28 Dec 2023


%\cite{LHCb:2019kea}
\bibitem{LHCb:2019kea}
R.~Aaij \textit{et al.} [LHCb],
%``Observation of a narrow pentaquark state, $P_c(4312)^+$, and of two-peak structure of the $P_c(4450)^+$,''
Phys. Rev. Lett. \textbf{122}, 222001 (2019).
%659 citations counted in INSPIRE as of 28 Dec 2023


%\cite{LHCb:2020jpq}
\bibitem{LHCb:2020jpq}
R.~Aaij \textit{et al.} [LHCb],
%``Evidence of a $J/\psi\Lambda$ structure and observation of excited $\Xi^-$ states in the $\Xi^-_b \to J/\psi\Lambda K^-$ decay,''
Sci. Bull. \textbf{66}, 1278-1287 (2021).
%176 citations counted in INSPIRE as of 28 Dec 2023


%\cite{LHCb:2022ogu}
\bibitem{LHCb:2022ogu}
R.~Aaij \textit{et al.} [LHCb],
%``Observation of a J/\ensuremath{\psi}\ensuremath{\Lambda} Resonance Consistent with a Strange Pentaquark Candidate in B-\textrightarrow{}J/\ensuremath{\psi}\ensuremath{\Lambda}p\textasciimacron{} Decays,''
Phys. Rev. Lett. \textbf{131}, 031901 (2023).
%64 citations counted in INSPIRE as of 28 Dec 2023



%\cite{BESIII:2026rzt}
\bibitem{BESIII:2026rzt}
M.~Ablikim \textit{et al.} [BESIII],
%``Observation of the $X(2370)$ in $J/¦×\rightarrow¦ÃK^{0}_{S}K^{0}_{S}¦Ð^{0}$ and $J/¦×\rightarrow¦Ã¦Ð^{0}¦Ð^{0}¦Ç$,''
[arXiv:2605.26495 [hep-ex]].
%3 citations counted in INSPIRE as of 31 Aug 2026


%\cite{Brambilla:2019esw}
\bibitem{Brambilla:2019esw}
N.~Brambilla, S.~Eidelman, C.~Hanhart, A.~Nefediev, C.~P.~Shen, C.~E.~Thomas, A.~Vairo and C.~Z.~Yuan,
%``The $XYZ$ states: experimental and theoretical status and perspectives,''
Phys. Rept. \textbf{873}, 1-154 (2020).
%722 citations counted in INSPIRE as of 27 Jun 2024


%\cite{Chen:2022asf}
\bibitem{Chen:2022asf}
H.~X.~Chen, W.~Chen, X.~Liu, Y.~R.~Liu and S.~L.~Zhu,
%``An updated review of the new hadron states,''
Rept. Prog. Phys. \textbf{86}, 026201 (2023).
%327 citations counted in INSPIRE as of 27 Jun 2024



%\cite{Meng:2022ozq}
\bibitem{Meng:2022ozq}
L.~Meng, B.~Wang, G.~J.~Wang and S.~L.~Zhu,
%``Chiral perturbation theory for heavy hadrons and chiral effective field theory for heavy hadronic molecules,''
Phys. Rept. \textbf{1019}, 1-149 (2023).
%138 citations counted in INSPIRE as of 27 Jun 2024


%\cite{Chen:2019bip}
\bibitem{Chen:2019bip}
H.~X.~Chen, W.~Chen and S.~L.~Zhu,
%``Possible interpretations of the $P_c(4312)$, $P_c(4440)$, and $P_c(4457)$,''
Phys. Rev. D \textbf{100}, 051501 (2019).
%143 citations counted in INSPIRE as of 28 Dec 2023

%\cite{Guo:2019fdo}
\bibitem{Guo:2019fdo}
F.~K.~Guo, H.~J.~Jing, U.~G.~Mei\ss{}ner and S.~Sakai,
%``Isospin breaking decays as a diagnosis of the hadronic molecular structure of the $P_c(4457)$,''
Phys. Rev. D \textbf{99},091501 (2019).
%117 citations counted in INSPIRE as of 28 Dec 2023


%\cite{Xiao:2019aya}
\bibitem{Xiao:2019aya}
C.~W.~Xiao, J.~Nieves and E.~Oset,
%``Heavy quark spin symmetric molecular states from ${\bar D}^{(*)}\Sigma_c^{(*)}$ and other coupled channels in the light of the recent LHCb pentaquarks,''
Phys. Rev. D \textbf{100},014021 (2019).
%148 citations counted in INSPIRE as of 28 Dec 2023


%\cite{He:2019ify}
\bibitem{He:2019ify}
J.~He,
%``Study of $P_c(4457)$, $P_c(4440)$, and $P_c(4312)$ in a quasipotential Bethe-Salpeter equation approach,''
Eur. Phys. J. C \textbf{79},393 (2019).
%149 citations counted in INSPIRE as of 28 Dec 2023


%\cite{Xiao:2019mvs}
\bibitem{Xiao:2019mvs}
C.~J.~Xiao, Y.~Huang, Y.~B.~Dong, L.~S.~Geng and D.~Y.~Chen,
%``Exploring the molecular scenario of Pc(4312) , Pc(4440) , and Pc(4457),''
Phys. Rev. D \textbf{100},014022 (2019).
%123 citations counted in INSPIRE as of 28 Dec 2023


%\cite{Roca:2015dva}
\bibitem{Roca:2015dva}
L.~Roca, J.~Nieves and E.~Oset,
%``LHCb pentaquark as a $\bar{D}^*\Sigma_c-\bar{D}^*\Sigma_c^*$ molecular state,''
Phys. Rev. D \textbf{92}, 094003 (2015).
%240 citations counted in INSPIRE as of 28 Dec 2023


%\cite{Chen:2015moa}
\bibitem{Chen:2015moa}
H.~X.~Chen, W.~Chen, X.~Liu, T.~G.~Steele and S.~L.~Zhu,
%``Towards exotic hidden-charm pentaquarks in QCD,''
Phys. Rev. Lett. \textbf{115},172001 (2015).
%243 citations counted in INSPIRE as of 28 Dec 2023


%\cite{Chen:2015loa}
\bibitem{Chen:2015loa}
R.~Chen, X.~Liu, X.~Q.~Li and S.~L.~Zhu,
%``Identifying exotic hidden-charm pentaquarks,''
Phys. Rev. Lett. \textbf{115},132002 (2015).
%276 citations counted in INSPIRE as of 28 Dec 2023


%\cite{Yang:2015bmv}
\bibitem{Yang:2015bmv}
G.~Yang and J.~Ping,
%``The structure of pentaquarks $P_c^+$ in the chiral quark model,''
Phys. Rev. D \textbf{95}, 014010 (2017).
%74 citations counted in INSPIRE as of 28 Dec 2023


%\cite{Huang:2015uda}25
\bibitem{Huang:2015uda}
H.~Huang, C.~Deng, J.~Ping and F.~Wang,
%``Possible pentaquarks with heavy quarks,''
Eur. Phys. J. C \textbf{76}, 624 (2016).
%105 citations counted in INSPIRE as of 28 Dec 2023


%\cite{Du:2019pij}
\bibitem{Du:2019pij}
M.~L.~Du, V.~Baru, F.~K.~Guo, C.~Hanhart, U.~G.~Mei\ss{}ner, J.~A.~Oller and Q.~Wang,
%``Interpretation of the LHCb $P_c$ States as Hadronic Molecules and Hints of a Narrow $P_c(4380)$,''
Phys. Rev. Lett. \textbf{124} (2020), 072001.
%140 citations counted in INSPIRE as of 04 Sep 2024


%\cite{BaBar:2003oey}
\bibitem{BaBar:2003oey}
B.~Aubert \textit{et al.} [BaBar],
%``Observation of a narrow meson decaying to $D_s^+ \pi^0$ at a mass of 2.32-GeV/c$^2$,''
Phys. Rev. Lett. \textbf{90} (2003), 242001.
%1009 citations counted in INSPIRE as of 25 Nov 2024


%\cite{CLEO:2003ggt}
\bibitem{CLEO:2003ggt}
D.~Besson \textit{et al.} [CLEO],
%``Observation of a narrow resonance of mass 2.46-GeV/c**2 decaying to D*+(s) pi0 and confirmation of the D*(sJ)(2317) state,''
Phys. Rev. D \textbf{68} (2003), 032002
[erratum: Phys. Rev. D \textbf{75} (2007), 119908].
%721 citations counted in INSPIRE as of 25 Nov 2024



%\cite{Belle:2003kup}
\bibitem{Belle:2003kup}
Y.~Mikami \textit{et al.} [Belle],
%``Measurements of the $D_{sJ}$ resonance properties,''
Phys. Rev. Lett. \textbf{92} (2004), 012002.
%299 citations counted in INSPIRE as of 25 Nov 2024



%\cite{Belle:2003guh}
\bibitem{Belle:2003guh}
P.~Krokovny \textit{et al.} [Belle],
%``Observation of the D(sJ)(2317) and D(sJ)(2457) in B decays,''
Phys. Rev. Lett. \textbf{91} (2003), 262002.
%399 citations counted in INSPIRE as of 25 Nov 2024



%\cite{BaBar:2004yux}
\bibitem{BaBar:2004yux}
B.~Aubert \textit{et al.} [BaBar],
%``Study of $B \to D_{sJ}^{(*)+} \bar{D}^{(*)}$ decays,''
Phys. Rev. Lett. \textbf{93} (2004), 181801.
%105 citations counted in INSPIRE as of 25 Nov 2024



%\cite{Xie:2010zza}
\bibitem{Xie:2010zza}
Z.~X.~Xie, G.~Q.~Feng and X.~H.~Guo,
%``Analyzing Ds0*(2317)(+) in the DK molecule picture in the Beth-Salpeter approach,''
Phys. Rev. D \textbf{81} (2010), 036014.
%36 citations counted in INSPIRE as of 25 Nov 2024


%\cite{Guo:2006fu}
\bibitem{Guo:2006fu}
F.~K.~Guo, P.~N.~Shen, H.~C.~Chiang, R.~G.~Ping and B.~S.~Zou,
%``Dynamically generated 0+ heavy mesons in a heavy chiral unitary approach,''
Phys. Lett. B \textbf{641} (2006), 278-285.
%285 citations counted in INSPIRE as of 25 Nov 2024



%\cite{Guo:2006rp}
\bibitem{Guo:2006rp}
F.~K.~Guo, P.~N.~Shen and H.~C.~Chiang,
%``Dynamically generated 1+ heavy mesons,''
Phys. Lett. B \textbf{647} (2007), 133-139.
%159 citations counted in INSPIRE as of 25 Nov 2024


%\cite{Gamermann:2006nm}
\bibitem{Gamermann:2006nm}
D.~Gamermann, E.~Oset, D.~Strottman and M.~J.~Vicente Vacas,
%``Dynamically generated open and hidden charm meson systems,''
Phys. Rev. D \textbf{76} (2007), 074016.
%340 citations counted in INSPIRE as of 25 Nov 2024


%\cite{Zhu:2019vnr}
\bibitem{Zhu:2019vnr}
H.~Zhu and Y.~Huang,
%``Production of the $D_{s0}(2317)$ and $D_{s1}(2460)$ by kaon-induced reactions on a proton target,''
Phys. Rev. D \textbf{100} (2019), 054031.
%1 citations counted in INSPIRE as of 25 Nov 2024


%\cite{Mohler:2013rwa}
\bibitem{Mohler:2013rwa}
D.~Mohler, C.~B.~Lang, L.~Leskovec, S.~Prelovsek and R.~M.~Woloshyn,
%``$D_{s0}^*(2317)$ Meson and $D$-Meson-Kaon Scattering from Lattice QCD,''
Phys. Rev. Lett. \textbf{111} (2013), 222001.
%205 citations counted in INSPIRE as of 25 Nov 2024


%\cite{Altenbuchinger:2013vwa}
\bibitem{Altenbuchinger:2013vwa}
M.~Altenbuchinger, L.~S.~Geng and W.~Weise,
%``Scattering lengths of Nambu-Goldstone bosons off $D$ mesons and dynamically generated heavy-light mesons,''
Phys. Rev. D \textbf{89} (2014), 014026.
%109 citations counted in INSPIRE as of 25 Nov 2024



%\cite{Hao:2022vwt}
\bibitem{Hao:2022vwt}
W.~Hao, Y.~Lu and B.~S.~Zou,
%``Coupled channel effects for the charmed-strange mesons,''
Phys. Rev. D \textbf{106} (2022), 074014.


%\cite{Jiang:2026rpu}
\bibitem{Jiang:2026rpu}
D.~Jiang, Y.~Huang and J.~Zhao,
%``Evidence for New $D_s$-Family Molecular States,''
[arXiv:2606.11717 [hep-ph]].
%1 citations counted in INSPIRE as of 01 Sep 2026


%\cite{Guo:2017jvc}
\bibitem{Guo:2017jvc}
F.~K.~Guo, C.~Hanhart, U.~G.~Mei{\ss}ner, Q.~Wang, Q.~Zhao and B.~S.~Zou,
%``Hadronic molecules,''
Rev. Mod. Phys. \textbf{90} (2018), 015004
[erratum: Rev. Mod. Phys. \textbf{94} (2022) no.2, 029901].


%\cite{Wang:2025jcq}
\bibitem{Wang:2025jcq}
L.~L.~Wang, X.~M.~Zhao, X.~H.~Liu and M.~J.~Yan,
%``New spectrum of charm-strange meson with constituent quark model \(c\overline{s}\) contributions,''
Phys. Lett. B \textbf{873} (2026), 140164.



%\cite{Guo:2011dd}
\bibitem{Guo:2011dd}
F.~K.~Guo and U.~G.~Meissner,
%``More kaonic bound states and a comprehensive interpretation of the $D_{sJ}$ states,''
Phys. Rev. D \textbf{84} (2011), 014013.


%\cite{Wang:2014jua}
\bibitem{Wang:2014jua}
Z.~G.~Wang,
%``$D_{s3}^*(2860)$ and $D_{s1}^*(2860)$ as the 1D $c\bar{s}$ states,''
Eur. Phys. J. C \textbf{75} (2015), 25.
%doi:10.1140/epjc/s10052-014-3246-z


%\cite{Li:2010vx}
\bibitem{Li:2010vx}
D.~M.~Li, P.~F.~Ji and B.~Ma,
%``The newly observed open-charm states in quark model,''
Eur. Phys. J. C \textbf{71} (2011), 1582.


%\cite{Zhang:2006yj}
\bibitem{Zhang:2006yj}
B.~Zhang, X.~Liu, W.~Z.~Deng and S.~L.~Zhu,
%``$D_ {sJ} (2860)$ and $D_ {sJ} (2715)$,''
Eur. Phys. J. C \textbf{50} (2007), 617-628.



%\cite{Close:2006gr}
\bibitem{Close:2006gr}
F.~E.~Close, C.~E.~Thomas, O.~Lakhina and E.~S.~Swanson,
%``Canonical interpretation of the D(sJ)(2860) and D(sJ)(2690),''
Phys. Lett. B \textbf{647} (2007), 159-163.


%\cite{Godfrey:2013aaa}
\bibitem{Godfrey:2013aaa}
S.~Godfrey and I.~T.~Jardine,
%``Nature of the $D_{s1}^*$(2710) and $D_{sJ}^*$(2860) mesons,''
Phys. Rev. D \textbf{89} (2014), 074023.


%\cite{Wang:2016hkf}
\bibitem{Wang:2016hkf}
Z.~G.~Wang,
%``Analysis of the strong decays $D_{s3}^*(2860)\to DK$, $D^{*}K$ with QCD sum rules,''
Eur. Phys. J. A \textbf{52} (2016), 303.
%doi:10.1140/epja/i2016-16303-y

%\cite{Jiang:2024ftx}
\bibitem{Jiang:2024ftx}
Z.~H.~Jiang and A.~Zhang,
%``Assignment of charmed-strange Ds0(2590)+ and DsJ(3040)+,''
Nucl. Phys. A \textbf{1048} (2024), 122893.


%\cite{Chen:2009zt}
\bibitem{Chen:2009zt}
B.~Chen, D.~X.~Wang and A.~Zhang,
%``Interpretation of D(sJ)(2632)+, D(s1)(2700)+-, D*(sJ)(2860)+ and D(sJ)(3040)+,''
Phys. Rev. D \textbf{80} (2009), 071502.


%\cite{Li:2009qu}
\bibitem{Li:2009qu}
D.~M.~Li and B.~Ma,
%``Implication of BaBar's new data on the D(s1)(2710) and D(sJ)(2860),''
Phys. Rev. D \textbf{81} (2010), 014021.


%\cite{Huang:2020ptc}
\bibitem{Huang:2020ptc}
Y.~Huang, J.~X.~Lu, J.~J.~Xie and L.~S.~Geng,
%``Strong decays of ${\bar{D}}^{*}K^{*}$ molecules and the newly observed $X_{0,1}$ states,''
Eur. Phys. J. C \textbf{80}, 973 (2020).



%\cite{Jafarzade:2021vhh}
\bibitem{Jafarzade:2021vhh}
S.~Jafarzade, A.~Koenigstein and F.~Giacosa,
%``Phenomenology of $J^{PC}$ = $3^{--}$ tensor mesons,''
Phys. Rev. D \textbf{103} (2021), 096027.


%\cite{Yang:2022rck}
\bibitem{Yang:2022rck}
F.~Yang, H.~Q.~Zhu and Y.~Huang,
%``Analysis of the {\ensuremath{\eta}}1(1855) as a KK{\textasciimacron}1(1400) molecular state,''
Nucl. Phys. A \textbf{1030} (2023), 122571.


%\cite{LHCb:2014ott}
\bibitem{LHCb:2014ott}
R.~Aaij \textit{et al.} [LHCb],
%``Observation of overlapping spin-1 and spin-3 $\bar{D}^0 K^-$ resonances at mass $2.86 {\rm GeV}/c^2$,''
Phys. Rev. Lett. \textbf{113} (2014), 162001.


%\cite{Guo:2022zbc}
\bibitem{Guo:2022zbc}
D.~Guo, J.~Z.~Wang, D.~Y.~Chen and X.~Liu,
%``Connection between near the $D_s^+D_s^-$ threshold enhancement in $B^+ \to D_s^+D_s^-K^+$ and conventional charmonium $\chi_{c0}(2P)$,''
Phys. Rev. D \textbf{106} (2022), 094037.



%\cite{Bando:1984ej}
\bibitem{Bando:1984ej}
M.~Bando, T.~Kugo, S.~Uehara, K.~Yamawaki and T.~Yanagida,
%``Is rho Meson a Dynamical Gauge Boson of Hidden Local Symmetry?,''
Phys. Rev. Lett. \textbf{54} (1985), 1215.


%\cite{Bando:1987br}
\bibitem{Bando:1987br}
M.~Bando, T.~Kugo and K.~Yamawaki,
%``Nonlinear Realization and Hidden Local Symmetries,''
Phys. Rept. \textbf{164} (1988), 217-314.


%\cite{Nagahiro:2008cv}
\bibitem{Nagahiro:2008cv}
H.~Nagahiro, L.~Roca, A.~Hosaka and E.~Oset,
%``Hidden gauge formalism for the radiative decays of axial-vector mesons,''
Phys. Rev. D \textbf{79} (2009), 014015.



%\cite{Yue:2024bvy}
\bibitem{Yue:2024bvy}
Z.~L.~Yue, Y.~Pan and D.~Y.~Chen,
%``Hidden charm decays of Y(4626) in a Ds*+Ds1(2536)- molecular frame,''
Phys. Rev. D \textbf{110} (2024), 074013.


%\cite{Guo:2016bjq}
\bibitem{Guo:2016bjq}
F.~K.~Guo, C.~Hanhart, Y.~S.~Kalashnikova, P.~Matuschek, R.~V.~Mizuk, A.~V.~Nefediev, Q.~Wang and J.~L.~Wynen,
%``Interplay of quark and meson degrees of freedom in near-threshold states: A practical parametrization for line shapes,''
Phys. Rev. D \textbf{93} (2016), 074031.












\end{thebibliography}
\end{document}